\documentclass[twocolumn]{emulateapj} 

\usepackage{amsmath}
\usepackage{amsfonts}
\usepackage{amssymb}
\usepackage{graphicx}
\usepackage{epstopdf}
\usepackage{epsfig}
\DeclareGraphicsExtensions{.pdf,.png,.jpg}
\usepackage{tabularx}
\usepackage{longtable}
\usepackage{rotating}
\usepackage{bold-extra}
\usepackage{appendix}
\usepackage{listings}
\usepackage{float}
\usepackage{natbib}

\def\Msun{M$_{\odot}$}
\def\Lsun{L$_{\odot}$}
\newcommand{\Cii}{[C\,{\scshape{ii}}]}
\newcommand{\CII}{[C\,{\small II}]}
\usepackage{url}
\usepackage{hyperref}
\usepackage{enumerate}
\newcommand{\comt}[1]{\ignorespaces}
\usepackage{morefloats}

\def\nstack{six}
\def\Nstack{Six}

\let\oldsqrt\sqrt
\def\sqrt{\mathpalette\DHLhksqrt}
\def\DHLhksqrt#1#2{%
\setbox0=\hbox{$#1\oldsqrt{#2\,}$}\dimen0=\ht0
\advance\dimen0-0.2\ht0
\setbox2=\hbox{\vrule height\ht0 depth -\dimen0}%
{\box0\lower0.4pt\box2}}
\shorttitle{A search for $z\sim4.5$ {\CII} emitters in AS2UDS}
\shortauthors{Cooke et al.}

\begin{document}


\title{An ALMA survey of the SCUBA-2 Cosmology Legacy Survey UKIDSS/UDS field: Identifying candidate \MakeLowercase{\emph{z}}\,$\sim4.5$ {\CII} emitters}


\author{E. A. Cooke$^{1}$}
\author{Ian Smail$^{1}$}
\author{A. M. Swinbank$^{1}$}
\author{S. M. Stach$^{1}$}
\author{Fang~Xia An$^{1,2}$}
\author{B. Gullberg$^{1}$}
\author{O. Almaini$^{3}$}
\author{C. J. Simpson$^{4}$}
\author{J. L. Wardlow$^{1}$}
\author{A. W. Blain$^{5}$}
\author{S. C. Chapman$^{6}$}
\author{Chian-Chou Chen$^{7}$}
\author{C. J. Conselice$^{3}$}
\author{K. E. K. Coppin$^{8}$}
\author{D. Farrah$^{9}$}
\author{D. T. Maltby$^{3}$}
\author{M. J. Micha{\l}owski$^{10}$}
\author{D. Scott$^{11}$}
\author{J. M. Simpson$^{12}$}
\author{A. P. Thomson$^{13}$}
\author{P. van der Werf$^{14}$}

\email{elizabeth.a.cooke@durham.ac.uk}
\affil{$^{1}$Centre for Extragalactic Astronomy, Department of Physics, Durham University, Durham, DH1 3LE, UK}
\affil{$^{2}$Purple Mountain Observatory, China Academy of Sciences, 2 West Beijing Road, Nanjing 210008, China}
\affil{$^{3}$School of Physics and Astronomy, University of Nottingham, University Park, Nottingham NG7 2RD, UK}
\affil{$^{4}$Gemini Observatory, Northern Operations Center, 670 N. A'oh{\=o}k{\=u} Place, Hilo HI 96720, USA}
\affil{$^{5}$Physics \& Astronomy, University of Leicester, 1 University Road, Leicester LE1 7RH, UK}
\affil{$^{6}$Department of Physics and Atmospheric Science, Dalhousie University, Halifax, NS B3H 4R2, Canada}
\affil{$^{7}$European Southern Observatory, Karl Schwarzschild Strasse 2, Garching, Germany}
\affil{$^{8}$Centre for Astrophysics Research, School of Physics, Astronomy and Mathematics, University of Hertfordshire, College Lane, Hatfield AL10 9AB, UK}
\affil{$^{9}$Department of Physics, Virginia Tech, Blacksburg, VA 24061, USA}
\affil{$^{10}$Astronomical Observatory Institute, Faculty of Physics, Adam Mickiewicz University, ul. S{\l}oneczna 36, 60-286 Pozna{\'n}, Poland}
\affil{$^{11}$Department of Physics \& Astronomy, University of British Columbia, 6224 Agricultural Road, Vancouver, BC V6T 1Z1, Canada}
\affil{$^{12}$Academia Sinica Institute of Astronomy and Astrophysics, No. 1, Sec. 4, Roosevelt Road, Taipei 10617, Taiwan}
\affil{$^{13}$Jodrell Bank Centre for Astrophysics, School of Physics and Astronomy, The University of Manchester, Oxford Road, Manchester, M13 9PL, UK}
\affil{$^{14}$Leiden Observatory, Leiden University, P.O. Box 9513, NL-2300 RA Leiden, The Netherlands}


\begin{abstract}
We report the results of a search for serendipitous {\CII}\,$157.74$\,\micron\ emitters at $z\simeq4.4$--$4.7$ using the Atacama Large Millimeter/submillimeter Array (ALMA). The search exploits the AS2UDS continuum survey, which covers $\sim50$\,arcmin$^2$ of the sky towards $695$ luminous ($S_{870}\gtrsim1$\,mJy) submillimeter galaxies (SMGs), selected from the SCUBA-2 Cosmology Legacy Survey (S2CLS) $0.96$\,deg$^2$ Ultra Deep Survey (UDS) field. 
We detect ten candidate line emitters, with an expected false detection rate of ten percent. All of these line emitters correspond to $870$\,\micron\ continuum-detected sources in AS2UDS. 
The emission lines in two emitters appear to be high-$J$\,CO, but the remainder have multi-wavelength properties consistent with \CII\ from $z\simeq4.5$ galaxies. 
Using our sample, we place a lower limit of $>5\times10^{-6}$\,Mpc$^{-3}$ on the space density of luminous ($L_{\rm IR} \simeq 10^{13}$\,\Lsun) SMGs at $z=4.40$--$4.66$, suggesting $\ge7$\,percent of SMGs with $S_{870\mu{\rm m}}\gtrsim1$\,mJy lie at $4<z<5$. 
From stacking the high-resolution ($\sim0.15''$ full-width half maximum) ALMA $870$\,\micron\ imaging, we show that the {\CII} line emission is more extended than the continuum dust emission, with an average effective radius for the \CII\ of $r_{\rm e} = 1.7^{+0.1}_{-0.2}$\,kpc compared to $r_{\rm e} = 1.0\pm0.1$\,kpc for the continuum (rest-frame $160$\,\micron). 
By fitting the far-infrared photometry for these galaxies from $100$--$870$\,\micron, we show that SMGs at $z\sim4.5$ have a median dust temperature of $T_{\rm d}=55\pm4$\,K. This is systematically warmer than $870$\,\micron-selected SMGs at $z\simeq2$, which typically have temperatures around $35$\,K. These $z\simeq4.5$ SMGs display a steeper trend in the luminosity-temperature plane than $z\le2$ SMGs. 
We discuss the implications of this result in terms of the selection biases of high redshift starbursts in far-infrared/submillimeter surveys. 
\end{abstract}


\section{Introduction}
Despite their high individual luminosities, ultra-luminous infrared galaxies (ULIRGs; $L_{\rm IR} > 10^{12}$\,\Lsun) contribute less than one percent of the local star-formation rate density \citep[e.g.,][]{Magnelli2011,Casey2012}. The situation at higher redshifts, however, appears to be very different. 
Measurements of the redshift distribution of high-redshift ULIRGs, including those detected at submillimeter wavelengths \citep[so-called ``submillimeter galaxies'', SMGs;][]{Smail1997} show a rapid rise ($\sim1000$-fold increase) in their volume density to a peak at $z\simeq2.5$ and a decline at high redshifts \citep[e.g.,][]{Aretxaga2003,Chapman2005,Wardlow2011,Yun2012,Casey2012,Simpson2014,Michalowski2017,Simpson2017a}. 
At $z\gtrsim1$ SMGs may contribute up to $50$\,percent of the star-formation rate density \citep[e.g.,][]{Peacock2000,Chapman2005,Barger2012,Swinbank2014,Casey2014,Zavala2017}. 
SMGs at higher redshifts ($z\gtrsim3$) may also hold the key to explaining the populations of $z\sim2$--$3$ compact, quiescent galaxies now being detected \citep[e.g.,][]{Toft2014,Hodge2016,Simpson2017b}. The high stellar masses and apparent old ages of these galaxies suggest that they formed in rapid, intense bursts of star formation at $z>3$ \citep[e.g.,][]{Glazebrook2017,Simpson2017b}. Such starbursts may be linked to high redshift SMGs, meaning these galaxies are an essential element in models of massive galaxy formation. 

High redshift ($z\gtrsim3$) SMGs therefore appear to play a potentially significant role in galaxy evolution, however their dusty nature and high redshift means that measuring their spectroscopic redshifts -- needed to constrain many of their basic properties -- is extremely challenging using ground-based optical/near-infrared spectroscopy. As a result, the redshift distribution of SMGs is increasingly incomplete at $z\gtrsim3$ \citep[e.g.,][]{Danielson2017}.  

Some progress can be made in identifying $z>3$ SMGs using far-infrared photometry to measure their infrared spectral energy distribution \citep[SED, e.g., to identify ``$500$\,\micron\ risers'';][]{Dowell2014}. However, the degeneracy in the SED shape between dust temperature and redshift make the derived redshifts highly uncertain \citep[e.g.,][]{Blain1999,Bethermin2015,Schreiber2018}. 

For the subset of optical/near-infrared bright SMGs where reliable photometric redshifts can be measured, recent studies have suggested that $z\gtrsim4$ SMGs are characterised by far-infrared SEDs that have warmer dust temperatures than SMGs at $z\simeq2$ \citep[$\sim40$--$50$\,K compared to $\sim35$\,K; e.g.,][]{Swinbank2014, Schreiber2017}.  
Although there are potentially biases in the derived characteristic temperatures due to selection effects, the higher dust temperatures inferred at $z\simeq4$ may also be driven by physical differences in galaxy properties compared to SMGs at $z\simeq2$, for example reflecting the size of the dust regions or the star-formation rate of the galaxy.  
Alternatively, higher star-formation efficiencies in $z\gtrsim4$ SMGs (which may have shorter dynamical times than SMGs at $z\simeq2$) may result in the warmer dust temperatures.  
However, these results rely on uncertain photometric redshifts and hence to reliably constrain any evolution in characteristic dust temperatures with redshift, precise spectroscopic redshifts for $z>3$ galaxies are required.

(Sub)millimeter spectroscopy provides one of the most reliable means to derive redshifts for distant SMGs, especially at $z\gtrsim3$ where the multi-wavelength counterparts are faint or undetected in the optical/near-infrared. 
With the advent of ALMA it is now possible to obtain high-resolution imaging and spectroscopy in submillimeter wavebands. This allows us to both efficiently target single-dish submillimeter sources and precisely locate the counterpart of the SMG, and also to search for emission lines in the far-infrared to measure spectroscopic redshifts.

The $^{2}$P$_{3/2} \rightarrow ^{2}$P$_{1/2}$ fine structure line of C$^+$ at $157.74$\,\micron\ (hereafter {\CII}) is one of the primary routes by which interstellar gas can cool and consequently is typically the strongest emission line in the far-infrared spectra of star-forming galaxies. \CII\ emission can account for up to two percent of the total bolometric luminosity in a galaxy \citep[e.g.,][]{Brauher2008}, although with one dex of scatter at a fixed far-infrared luminosity \citep[e.g.,][]{DiazSantos2013}. 
The scatter arises due to the complex mix of processes that generate \CII\ emission. For example, 
{\CII} emission can originate both in photodissociation regions around star-forming regions and also from atomic and ionized gas \citep[e.g.,][]{Dalgarno1972,Madden1997,Pineda2013}. \CII\ could thus provide information about the volume and extent of the cold gas reservoir and star formation in galaxies. 
In particular for star-forming galaxies the photodissociation regions can dominate the \CII\ emission and so several studies have shown the {\CII} emission line correlates with the star-formation rate \citep[e.g.,][]{Stacey1991,GraciaCarpio2011,DeLooze2014}.

To date, one of the largest samples of interferometrically-identified submillimeter galaxies available is the ALMA-LABOCA Extended \emph{Chandra} Deep Field-South Survey \citep[ALESS;][]{Hodge2013}, which identified $99$ SMGs, $21$ of which are likely to lie at $z>4$ given their multi-wavelength properties \citep{Simpson2014}. 
At $z\sim4$--$5$ {\CII} is redshifted to $\sim870$\,\micron. 
In two of the ALESS sources {\CII} was serendipitously detected in the ALMA Band 7 observations at a redshift of $z=4.42$--$4.44$ \citep{Swinbank2012}, placing weak constraints on the properties of these galaxies and the {\CII} luminosity function at this redshift. 
However, with only two sources, a larger spectroscopic sample is clearly required in order to improve our understanding of the properties of $z\ge4$ SMGs. 

To increase the sample size of high-redshift SMGs, we have undertaken the  ALMA-SCUBA-2 survey of the UDS field (AS2UDS): an ALMA Band 7 survey of all $716$ submillimeter sources detected in the UKIDSS Ultra Deep Survey field by SCUBA-2 on the James Clerk Maxwell Telescope \citep[JCMT;][]{Geach2017}. This survey has precisely located $695$ SMGs (Stach et al., in preparation). Here we examine the ALMA datacubes to search for serendipitous emission lines. The frequency coverage of our data is $336$--$340$ and $348$--$352$\,GHz, corresponding to $z=4.40$--$4.46$ and $z=4.60$--$4.66$ for {\CII}. We aim to spectroscopically confirm \CII\ emission line sources at $z\simeq4.5$ and thus determine their basic properties such as infrared luminosity and dust temperature, as well as measure the number density of SMGs at $z>4$. 

The paper is laid out as follows: in Section \ref{sec:data} we outline the observations and data reduction. Section \ref{sec:results} presents our results and discussion. Our conclusions are given in Section \ref{sec:conclusions}. Throughout we use AB magnitudes and assume a $\Lambda$CDM cosmology with $\Omega_{\rm M}=0.3$, $\Omega_\Lambda=0.7$ and $H_{0}=70$\,km\,s$^{-1}$\,Mpc$^{-1}$. 

\begin{figure*}         
\includegraphics[trim= 8.0cm  6.9cm  1.0cm  0.0cm,clip,width=0.224\textwidth]{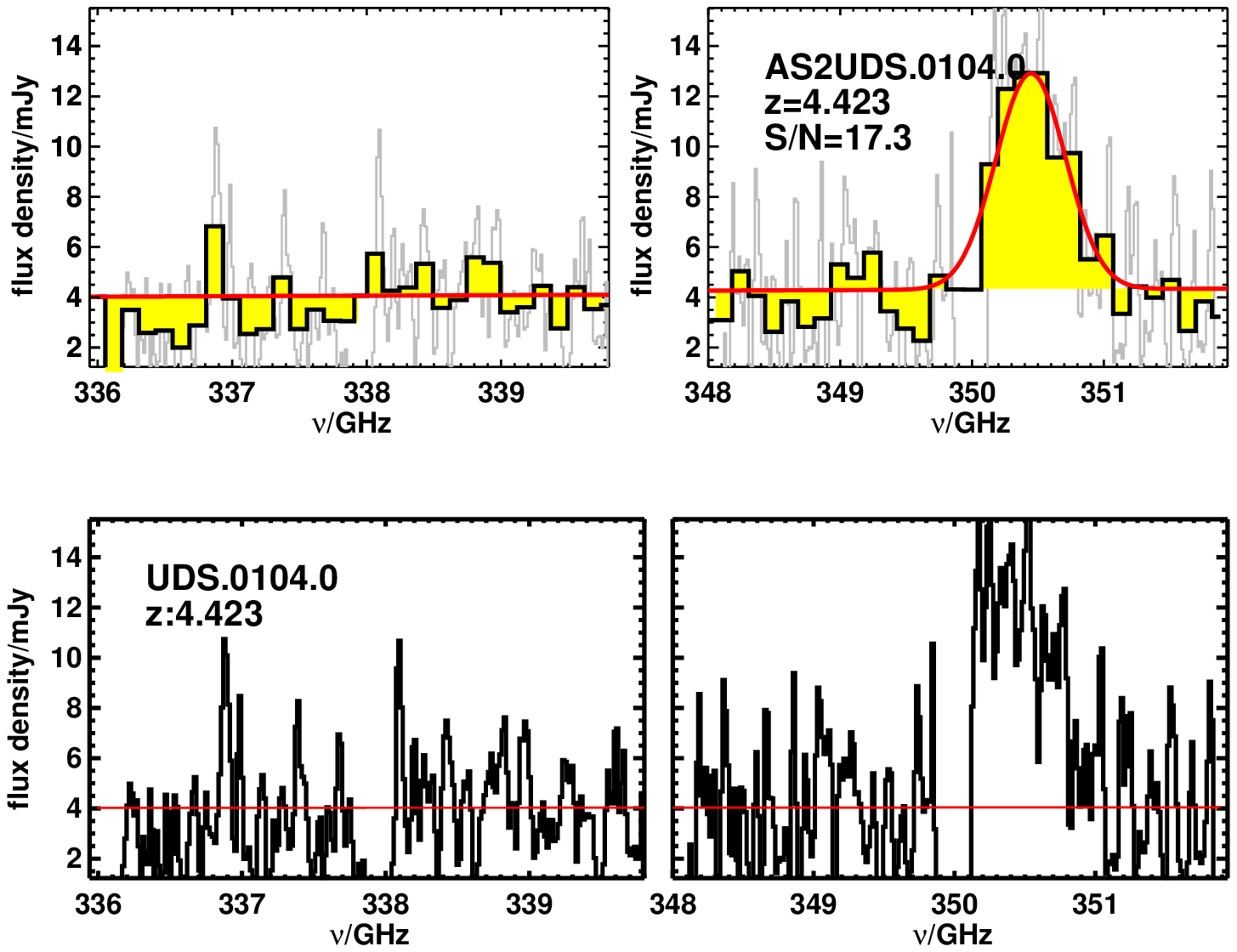}
\hspace{-0.54cm} \noindent
\includegraphics[trim= 8.5cm  6.9cm  1.0cm  0.0cm,clip,width=0.210\textwidth]{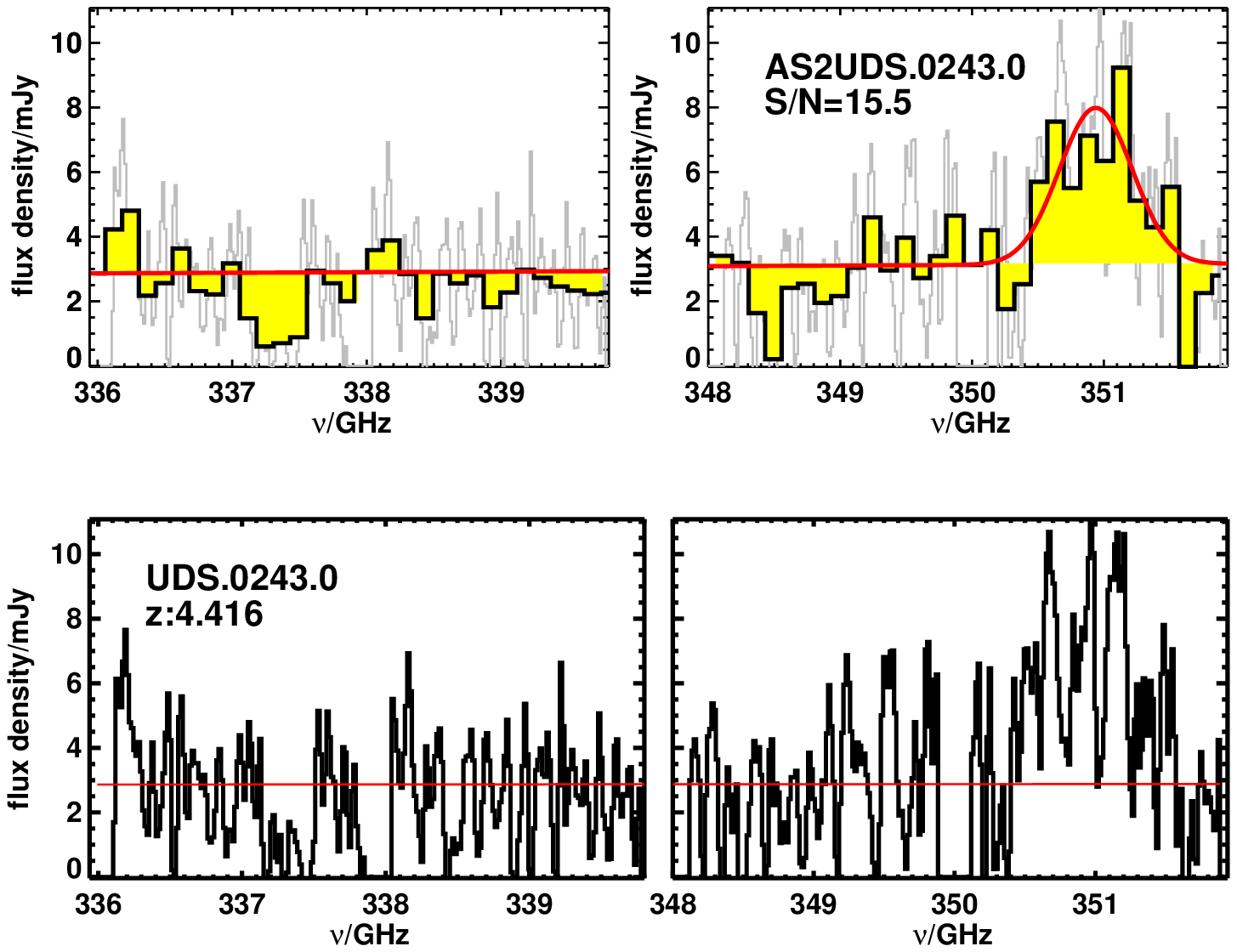}
\hspace{-0.54cm} \noindent
\includegraphics[trim= 1.1cm  6.9cm  8.9cm  0.0cm,clip,width=0.196\textwidth]{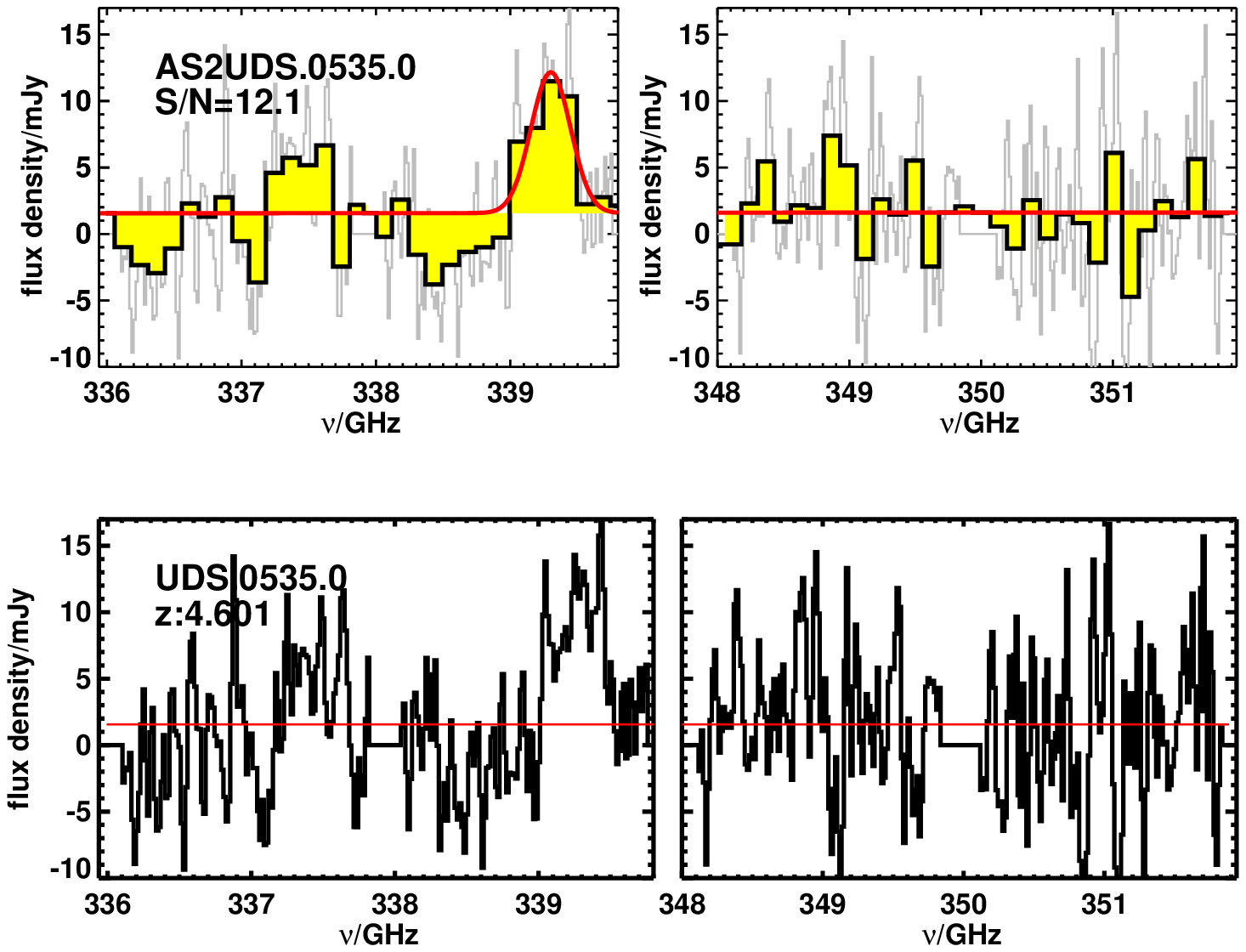}
\hspace{-0.29cm} \noindent
\includegraphics[trim= 8.5cm  6.9cm  1.0cm  0.0cm,clip,width=0.210\textwidth]{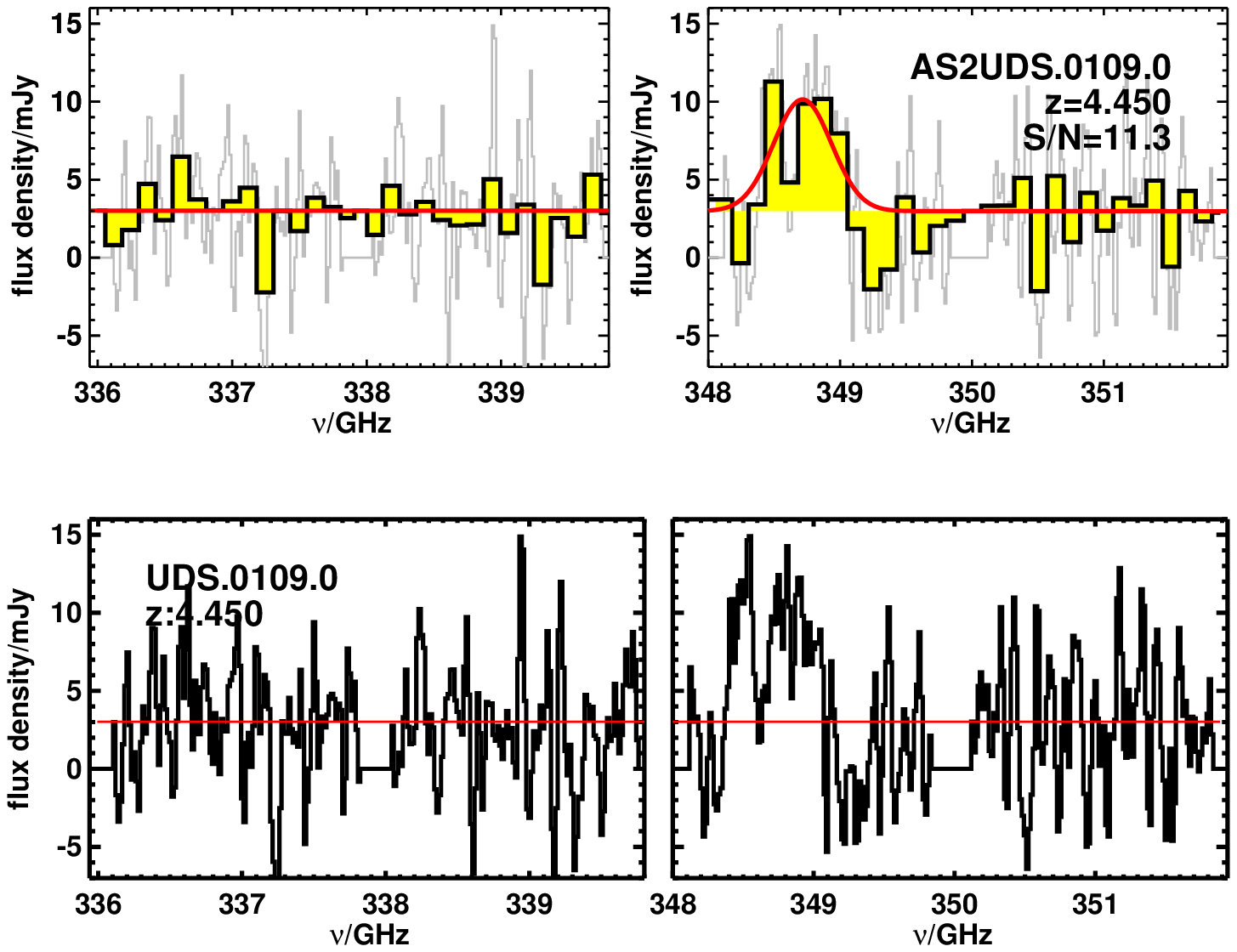}
\hspace{-0.54cm} \noindent
\includegraphics[trim= 8.5cm  6.9cm  1.0cm  0.0cm,clip,width=0.210\textwidth]{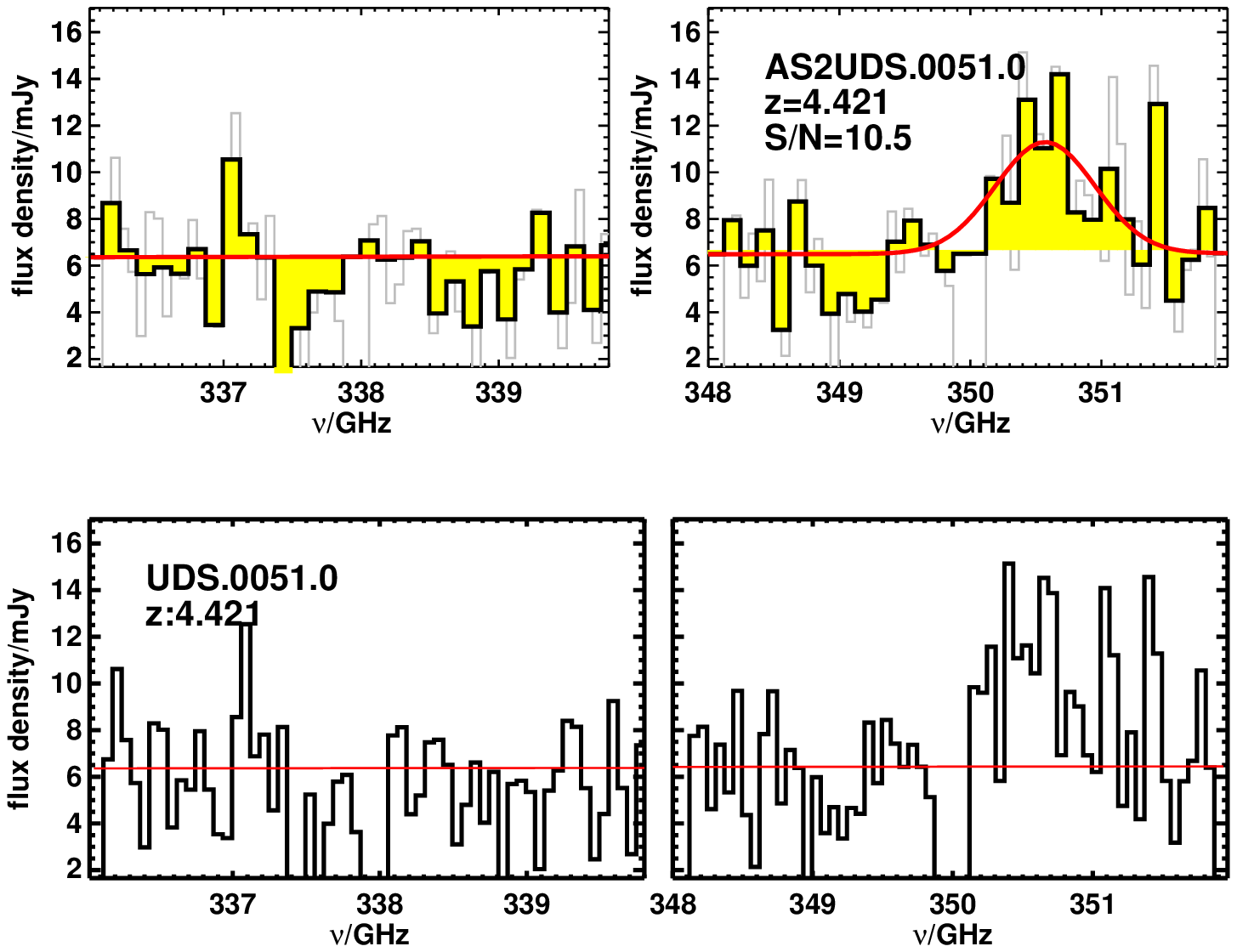}
\hspace{-0.54cm} \noindent

\vspace{-0.6cm} \noindent
\includegraphics[trim= 8.0cm  6.5cm  1.0cm  0.0cm,clip,width=0.224\textwidth]{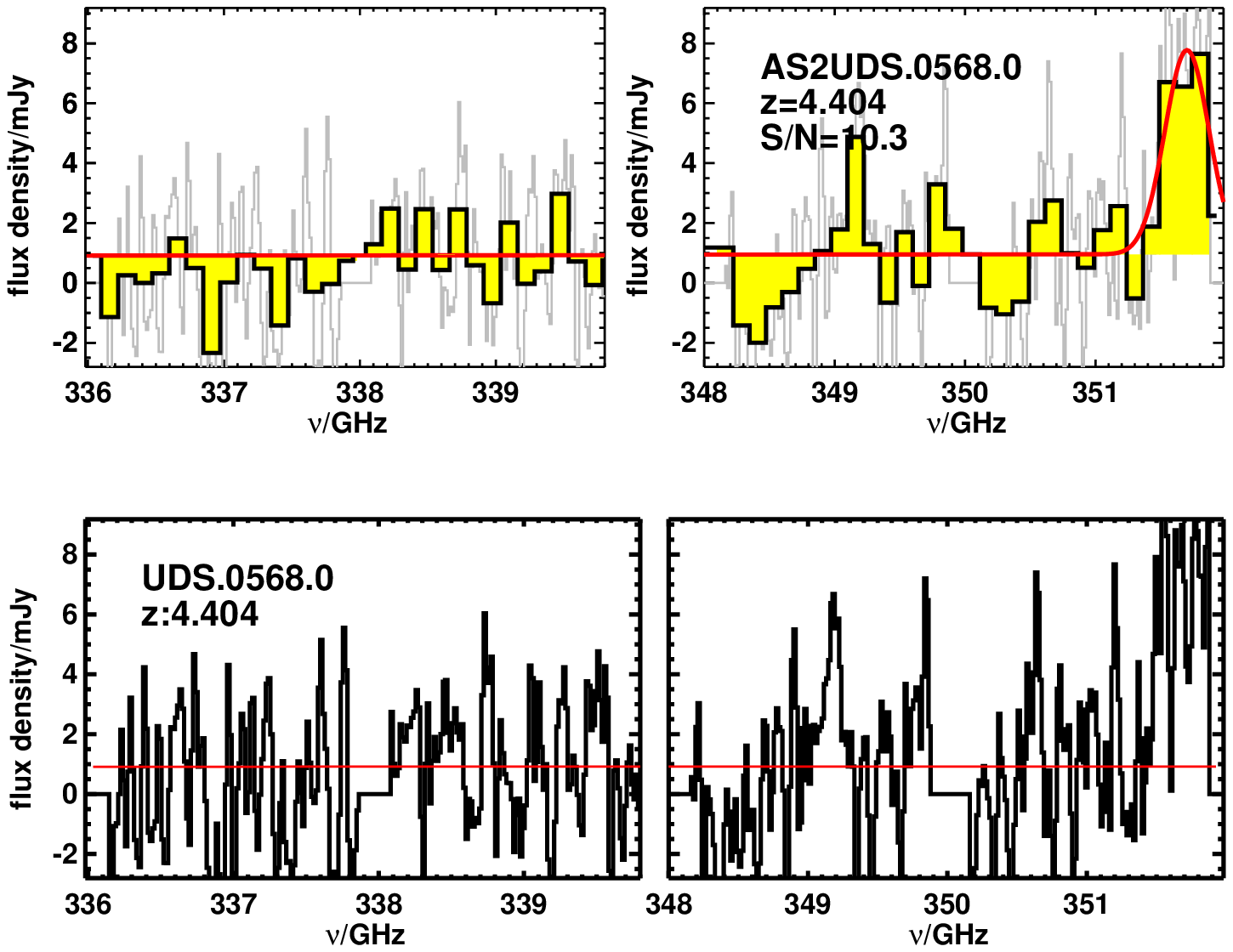}
\hspace{-0.54cm} \noindent
\includegraphics[trim= 1.1cm  6.5cm  8.9cm  0.0cm,clip,width=0.196\textwidth]{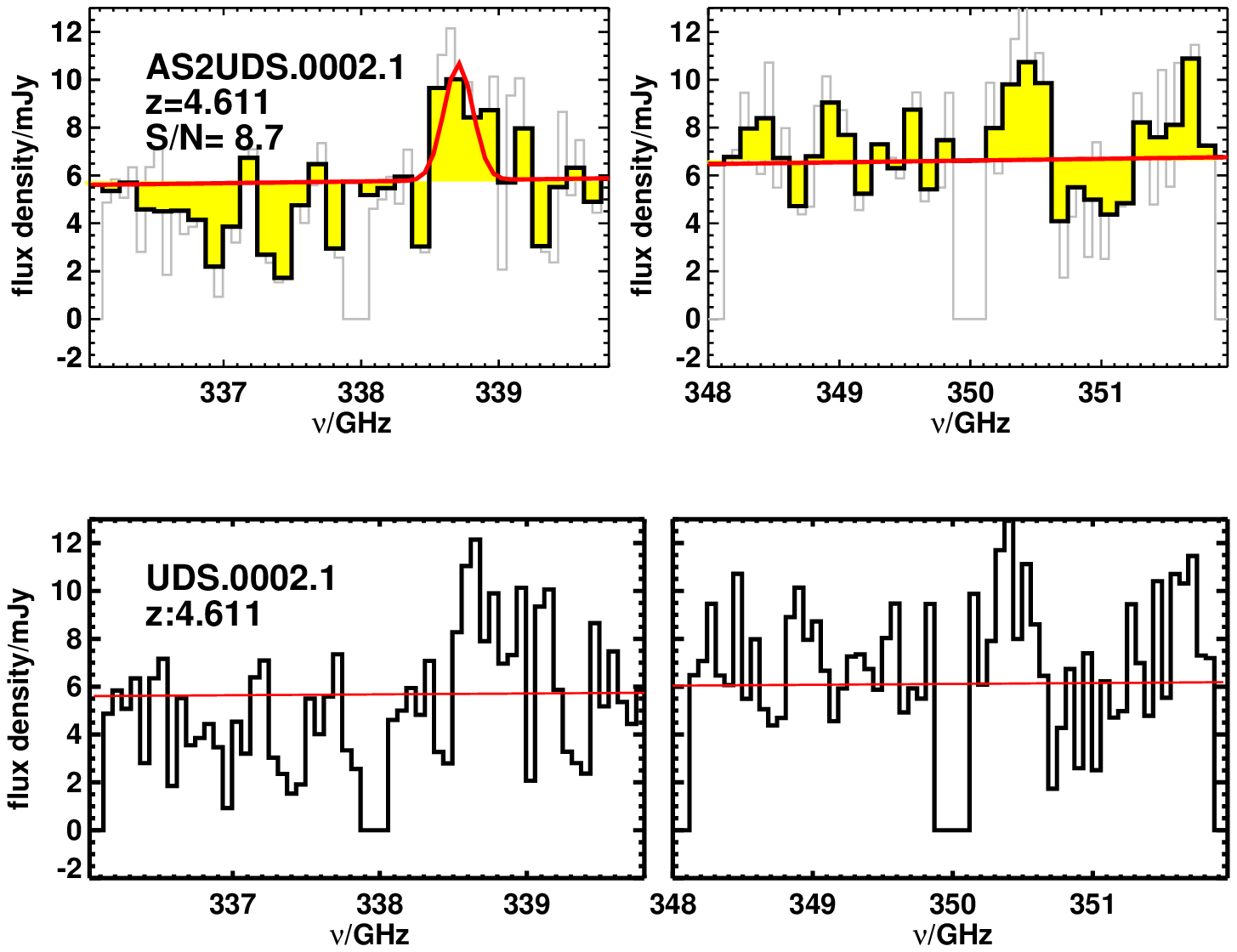}
\hspace{-0.29cm} \noindent
\includegraphics[trim= 1.1cm  6.5cm  8.9cm  0.0cm,clip,width=0.196\textwidth]{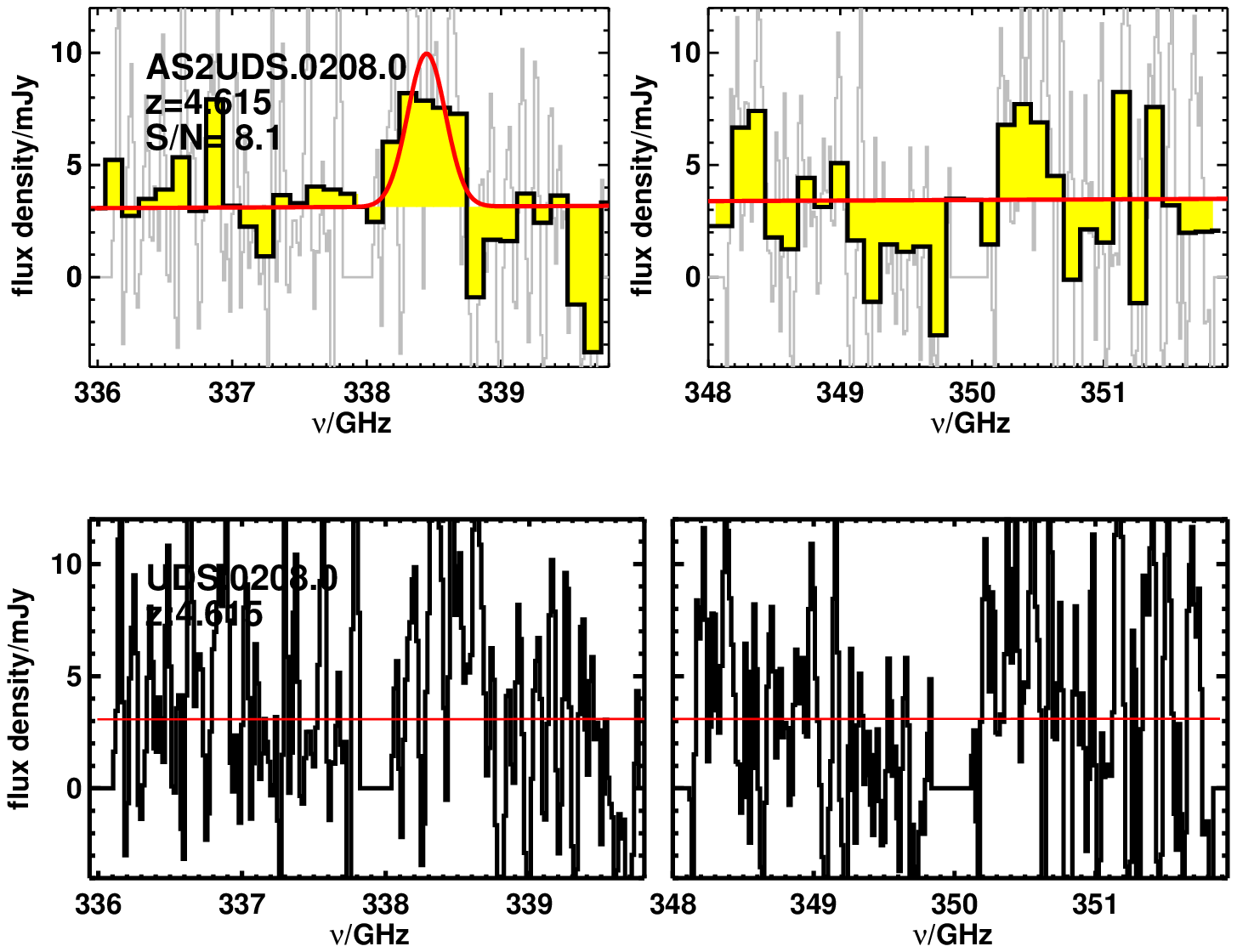}
\hspace{-0.29cm} \noindent
\includegraphics[trim= 8.5cm  6.5cm  1.0cm  0.0cm,clip,width=0.210\textwidth]{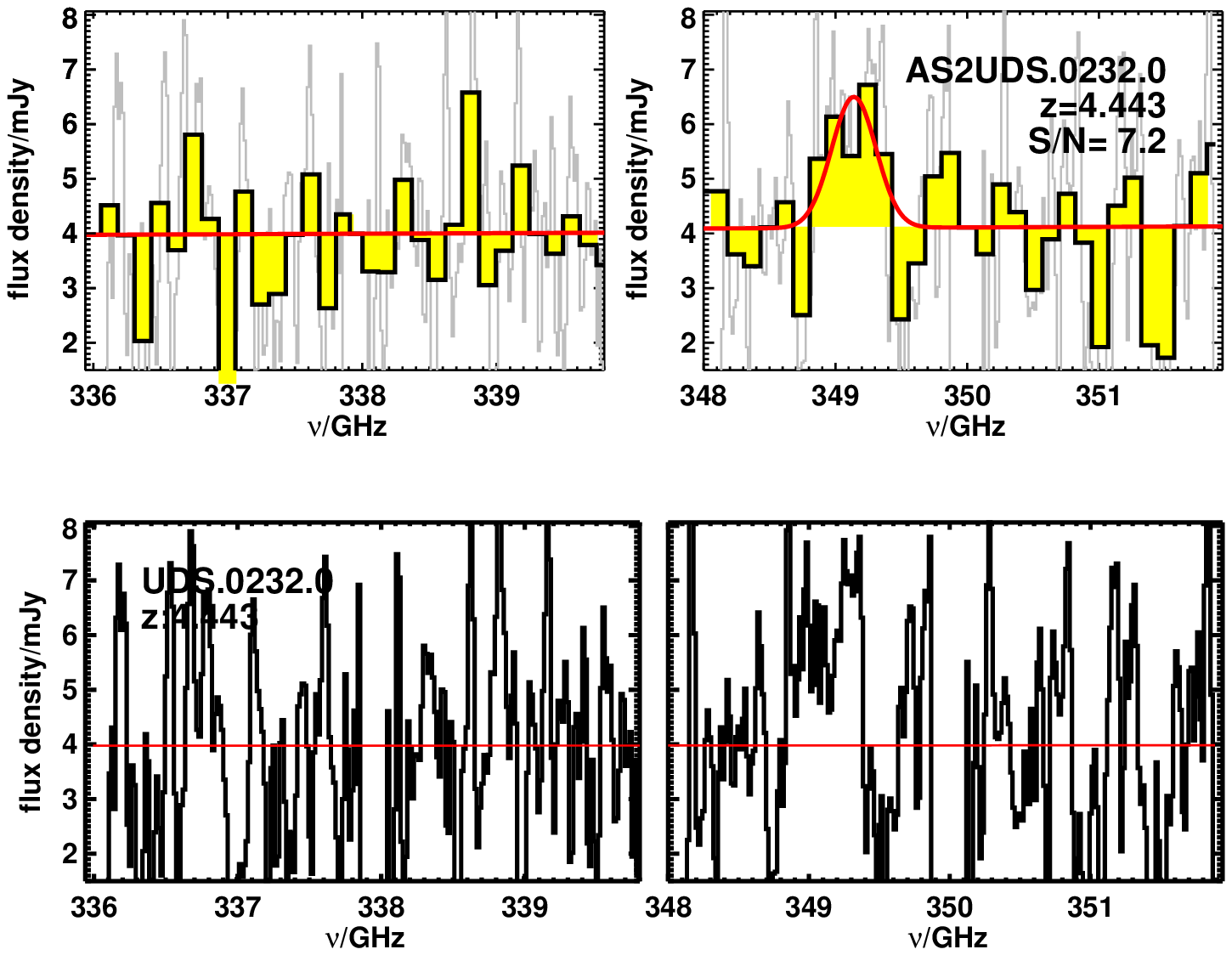}
\hspace{-0.54cm} \noindent
\includegraphics[trim= 1.1cm  6.5cm  8.9cm  0.0cm,clip,width=0.196\textwidth]{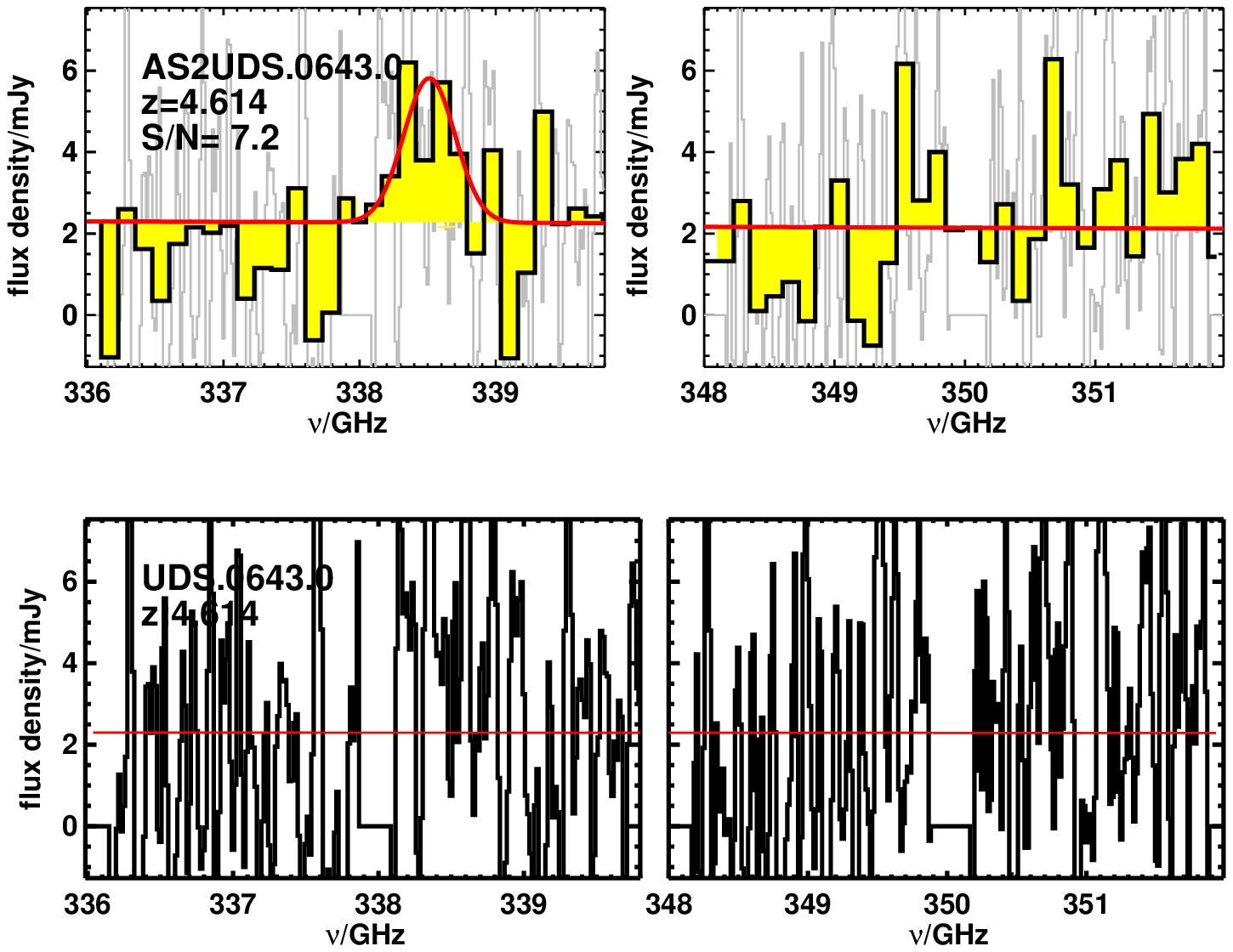}
\hspace{-0.29cm} \noindent

\caption{
Emission lines detected in the AS2UDS Band 7 datacubes, ranked by their integrated signal-to-noise. These are labelled with the SMG ID, the signal-to-noise and the corresponding redshift if the line is {\Cii}\,157.74\,\micron. 
We find ten line emitters with integrated signal-to-noise ranging from ${\rm S/N}=7.2$--$17.3$ in the $695$ SMGs with $S_{870\mu{\rm m}}\gtrsim1$\,mJy. 
Fluxes shown are not primary beam-corrected (typically a correction of $<10$\,percent) and are measured at the position of peak flux within the $0.5$\,arcsec tapered dirty cubes. 
The grey lines show the unbinned data. The black histogram shows the data binned to 100\,km\,s$^{-1}$. The red lines show the Gaussian fit to the detected emission line and continuum level. 
We note that AS2UDS.0243.0 and AS2UDS.0535.0 have photometric properties which suggest the line we detect is CO(8--7) or CO(5--4) respectively, corresponding to $z_{\rm CO} < 2$. Further observations are needed to confirm the nature of these emission lines. }
\label{fig:spectra}
\end{figure*}

\begin{figure*}  
\includegraphics[width=0.2\textwidth, trim={0.2cm 0.9cm 0.5cm 1cm}, clip]{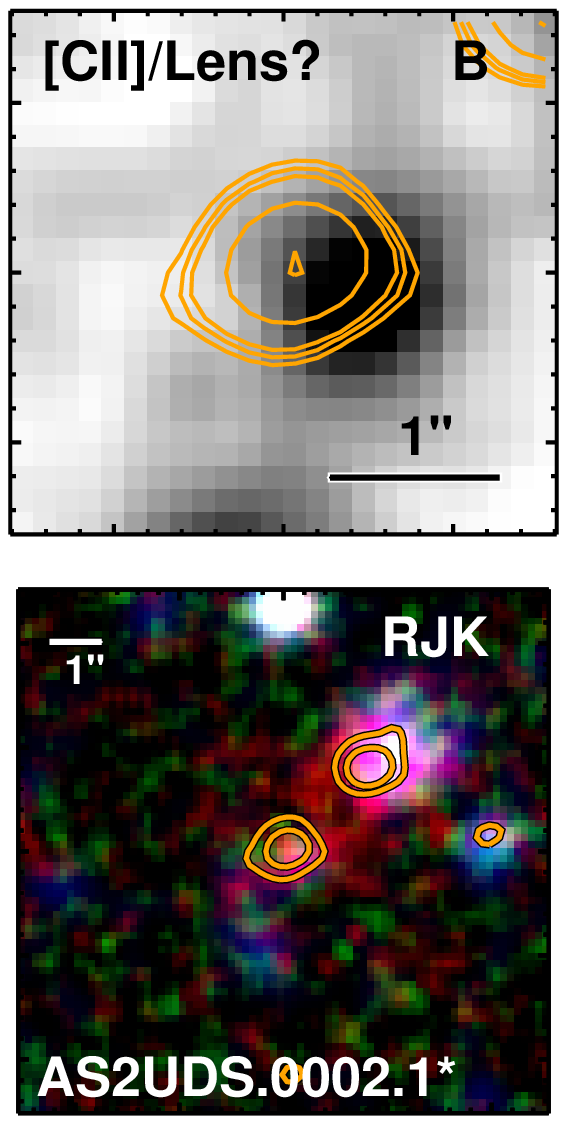}
\hspace{-0.45cm} \noindent   
\includegraphics[width=0.2\textwidth, trim={0.2cm 0.9cm 0.5cm 1cm}, clip]{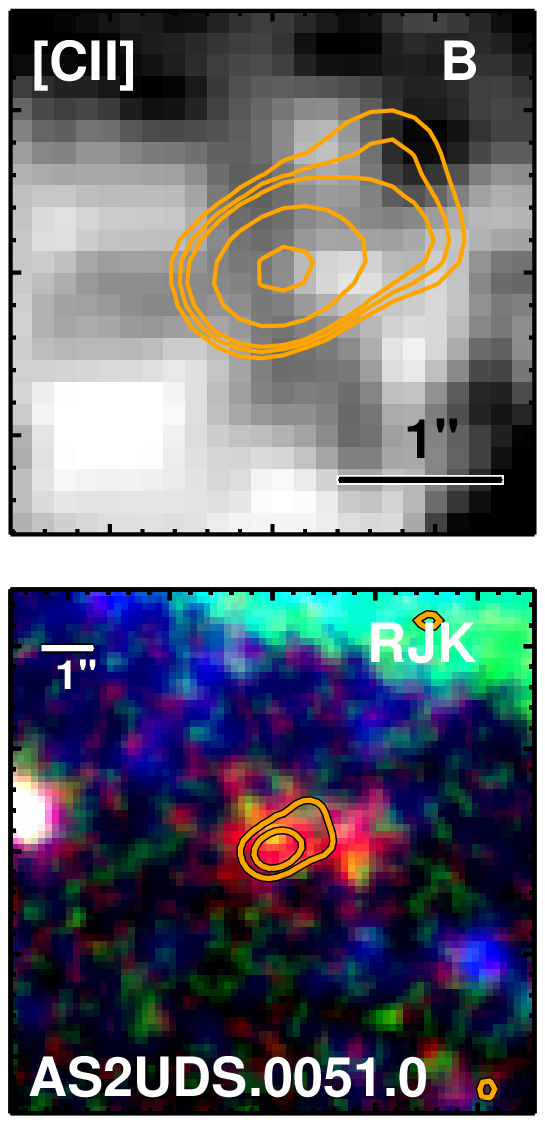}
\hspace{-0.45cm} \noindent     
\includegraphics[width=0.2\textwidth, trim={0.2cm 0.9cm 0.5cm 1cm}, clip]{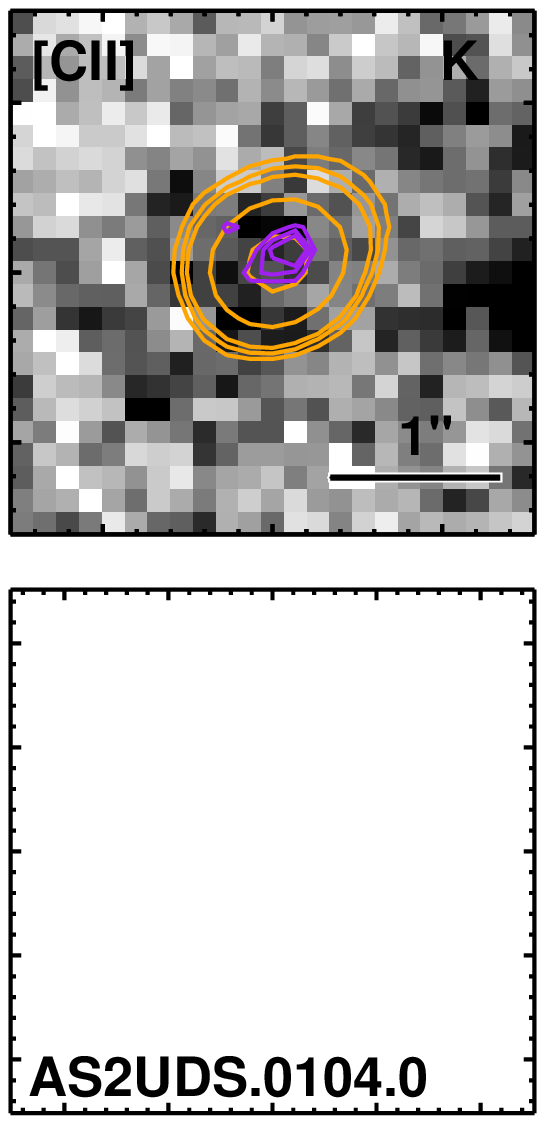}
\hspace{-0.45cm} \noindent 
\includegraphics[width=0.2\textwidth, trim={0.2cm 0.9cm 0.5cm 1cm}, clip]{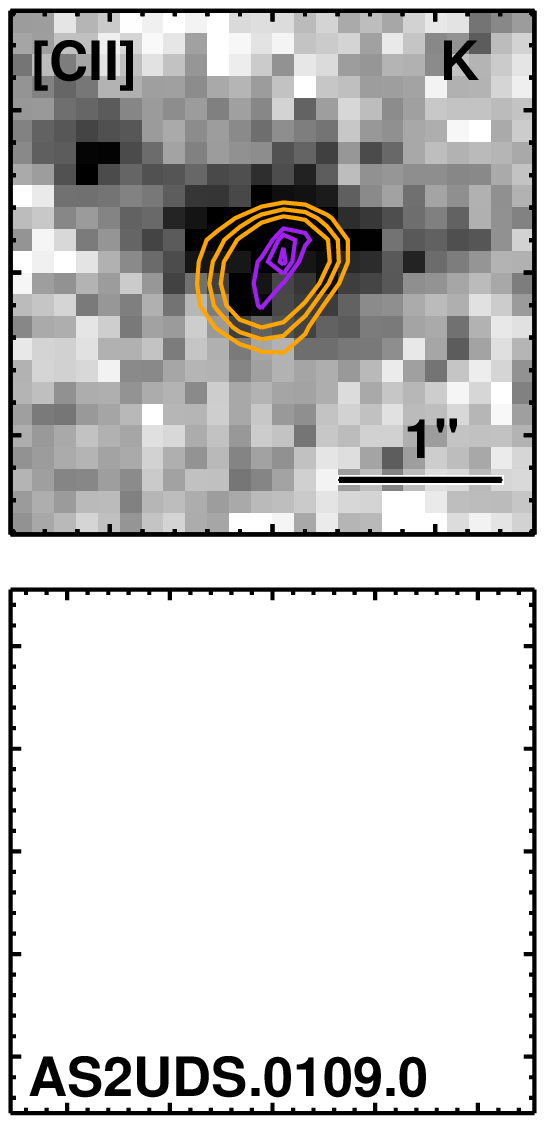}  
\hspace{-0.45cm} \noindent     
\includegraphics[width=0.2\textwidth, trim={0.2cm 0.9cm 0.5cm 1cm}, clip]{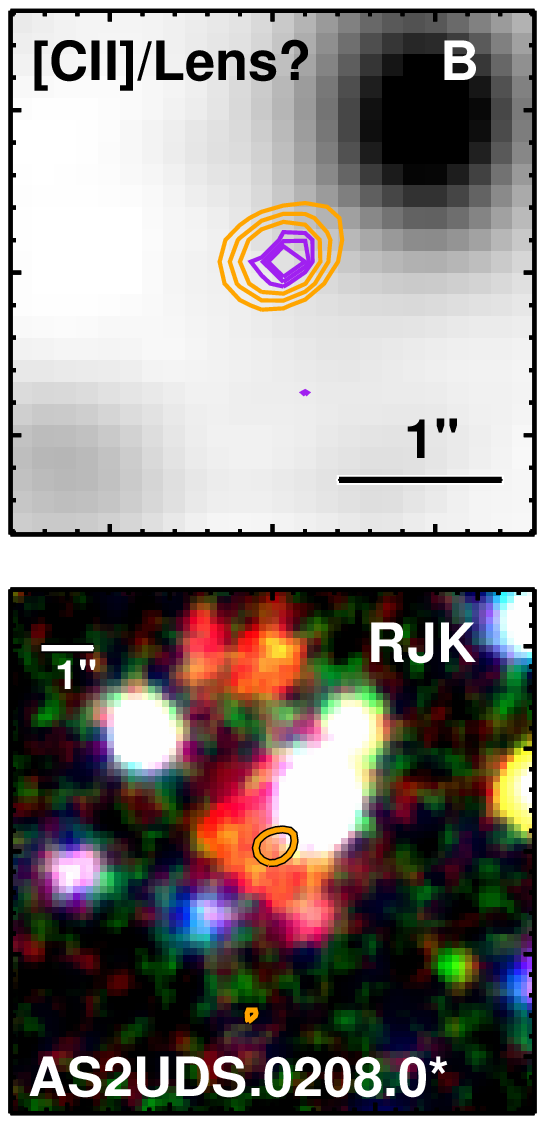}

\vspace{0.6cm} \noindent
\includegraphics[width=0.2\textwidth, trim={0.2cm 0.9cm 0.5cm 1cm}, clip]{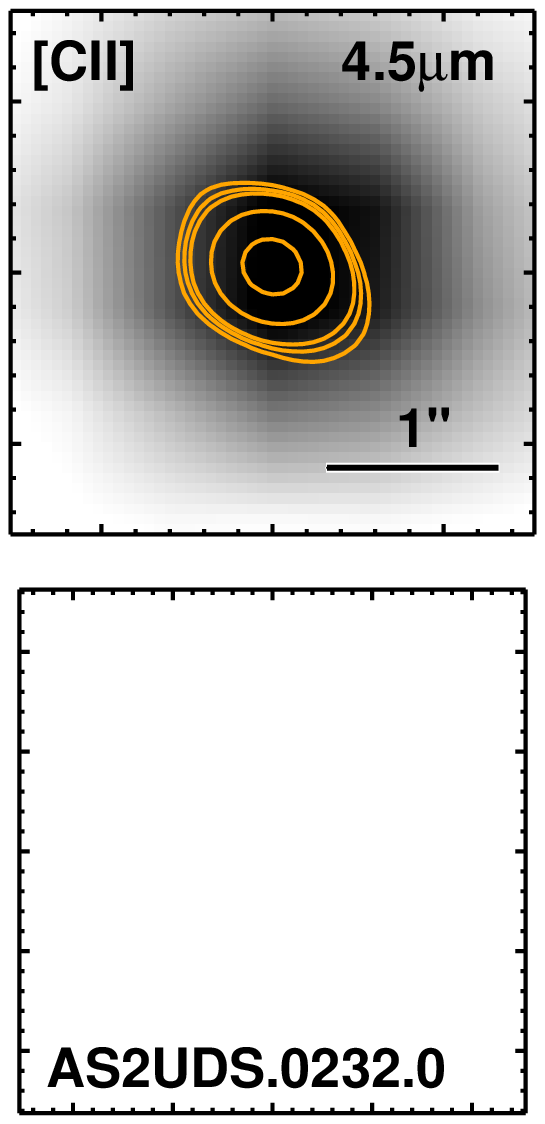}
\hspace{-0.45cm} \noindent 
\includegraphics[width=0.2\textwidth, trim={0.2cm 0.9cm 0.5cm 1cm}, clip]{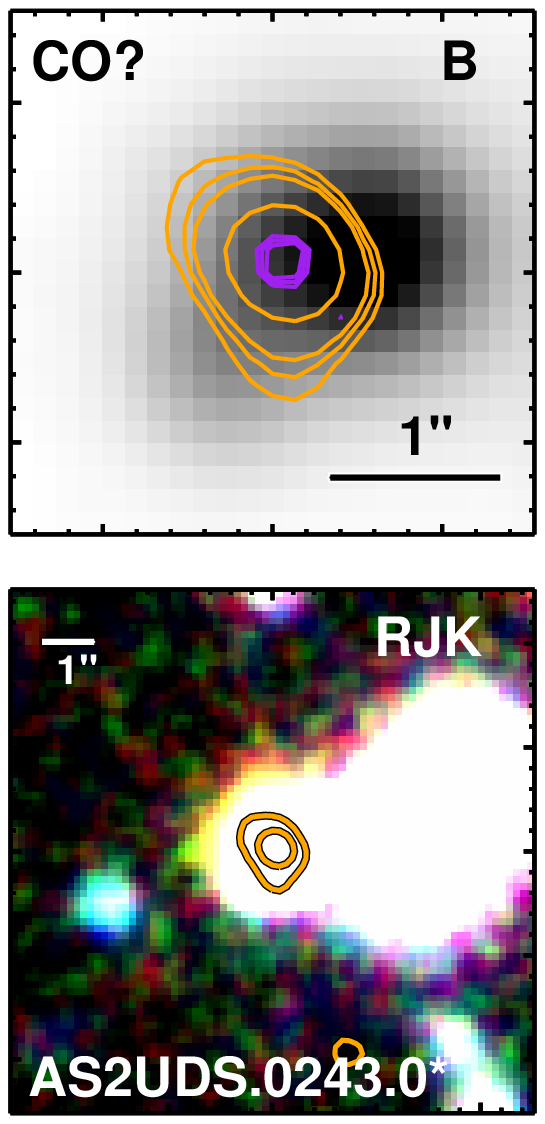}
\hspace{-0.45cm} \noindent  
\includegraphics[width=0.2\textwidth, trim={0.2cm 0.9cm 0.5cm 1cm}, clip]{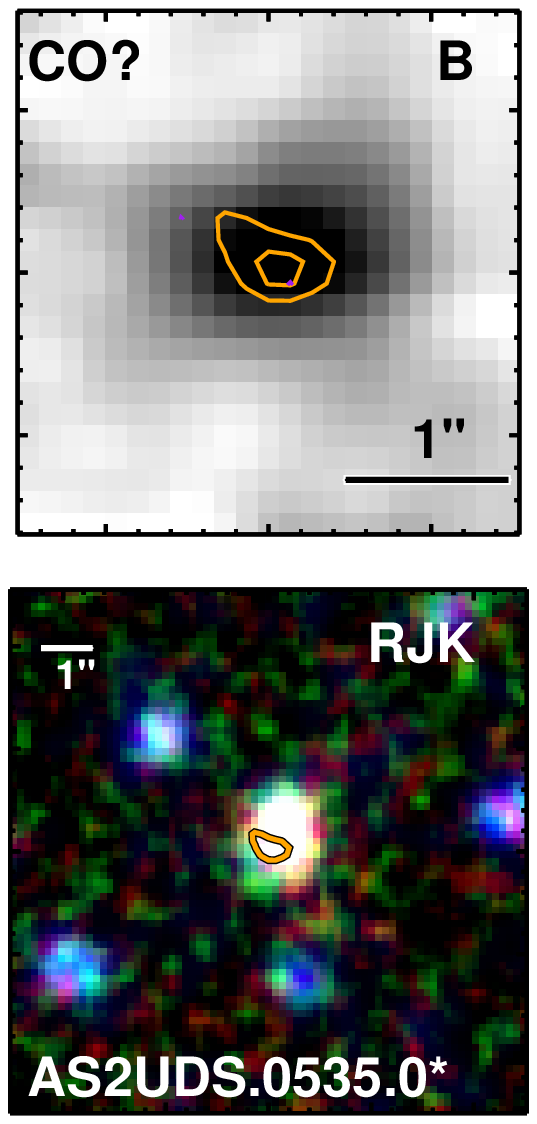}
\hspace{-0.45cm} \noindent     
\includegraphics[width=0.2\textwidth, trim={0.2cm 0.9cm 0.5cm 1cm}, clip]{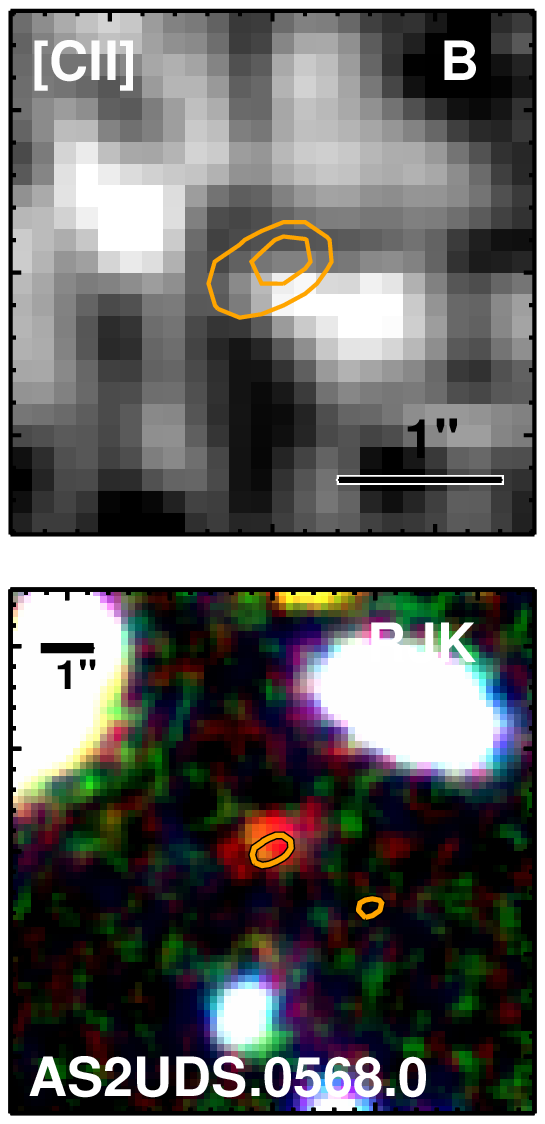}
\hspace{-0.45cm} \noindent     
\includegraphics[width=0.2\textwidth, trim={0.2cm 0.9cm 0.5cm 1cm}, clip]{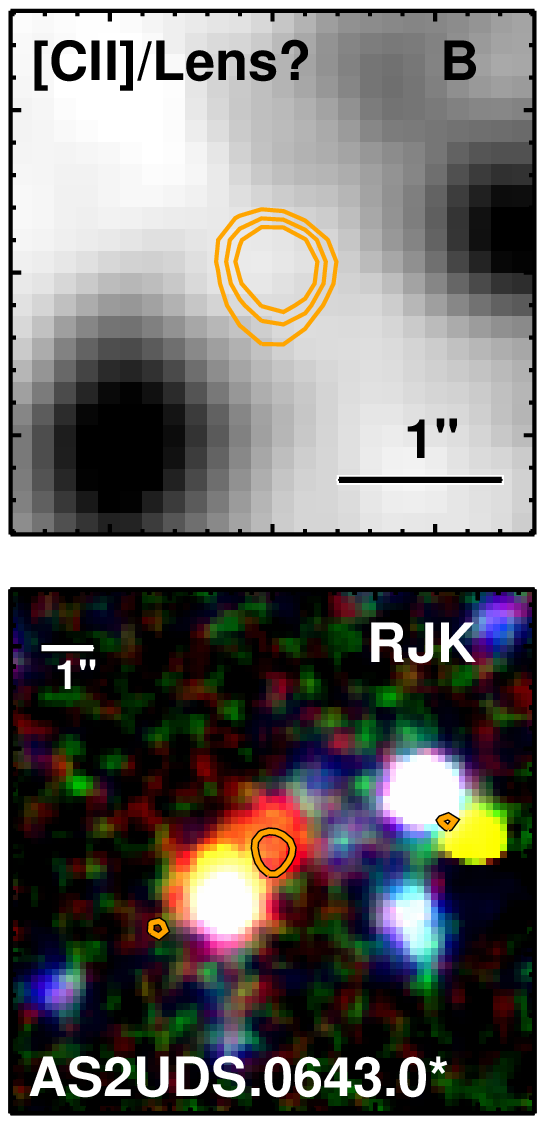}

\caption{Thumbnails of the line emitters detected in our survey. 
Top row in each panel: $3''\times3''$ thumbnails with ALMA continuum contours overlaid at $3, 4, 5, 10,$ and $20\sigma$ on background images. These show $B$-band unless the $B$-band is not available, in which case they show the $K$ or IRAC $4.5$\,\micron\ bands. 
Purple contours are high-resolution ALMA continuum maps ($\sim0.15''$) where it is available, orange are at the tapered $0.5''$ resolution. 
Bottom row in each panel: $10''\times10''$ true colour $R,J,K$ thumbnails of the line emitters where available. All thumbnails are centred on the tapered ALMA continuum emission, shown by the orange contours at $3$ and $10\sigma$. 
We label each pair of panels with the source ID and our classification of the source (see Table \ref{table:photometry}), asterisks indicate that the photometry of the ALMA source may be contaminated by a foreground galaxy, lensed or that the galaxy actually lies at $z<4$ and the detected line is not {\Cii} (see discussion in Section \ref{sec:COvCII}). 
}
\label{fig:thumbs}
\end{figure*}

\section{Observations, data reduction and analysis} \label{sec:data}
\subsection{ALMA data} \label{sec:almadata}
The UDS $0.96$\,degree$^2$ field was observed at $850$\,\micron\ with SCUBA-2 as part of the Cosmology Legacy Survey \citep{Geach2017} to a depth of $\sigma_{850}\simeq0.9$\,mJy\,beam$^{-1}$, detecting $716$ submillimeter sources above $4\sigma$, $S_{850}\simeq 3.6$\,mJy . We observed all $716$ of these submillimeter sources with ALMA at $870$\,\micron\ in Band 7 to pinpoint the galaxies responsible for the submillimeter emission. The data were taken in the period 2013 November to 2017 May (Cycles $1$, $3$, and $4$) with a dual polarization set up.  
The full data reduction and catalogue will be presented in Stach et al.\ (in preparation). In brief, our observations cover a total bandwidth of 7.5\,GHz split into two sidebands: $336$--$340$\,GHz and $348$--$352$\,GHz. The synthesised beam of our observations is $0.15$--$0.3$\,arcsec full-width, half maximum (FWHM), adopting natural weighting. The primary beam of ALMA is $\sim18$\,arcsec FWHM, which covers the SCUBA-2 beam (FWHM$\sim14.5$\,arcsec). This coverage, combined with the higher resolution and greater depth of the ALMA observations, means that we expect to detect the sources responsible for the original SCUBA-2 detections in the ALMA continuum data. 

Each pointing was centred on the SCUBA-2 catalogue position and observed for a total of $\sim40$\,s. A subset of $120$ of the pointings were observed in both Cycle $3$ and $4$ and thus have a longer total integration time (typically $80$--$90$\,s). 
All data were processed using the Common Astronomy Software Application \citep[{\sc{casa}};][]{McMullin2007}. 
We construct cleaned, tapered continuum maps and dirty, tapered data cubes. For more reliable line detections, we image the cubes without applying any cleaning to deconvolve the beam. 
The final cleaned, $0.5''$ FWHM-tapered continuum maps have average depths of $\sigma_{870\mu{\rm m}}=0.25,0.34$, and $0.23$\,mJy\,beam$^{-1}$ for Cycle $1,3$, and $4$ data, respectively. 
Stach et al.\ (in preparation) present an analysis of the data, catalogue construction and multi-wavelength properties. 

ALMA continuum sources were identified in the continuum maps as submillimeter sources with a signal-to-noise ratio ${\rm S/N}\ge4.3$ within a $0.5$\,arcsec diameter aperture, calculated from the aperture-integrated flux and the noise measured in randomly-placed apertures for each map. This S/N limit provides a two percent false positive rate, as determined by inverting the maps. In total we selected $695$ ALMA continuum-detected SMGs brighter than $S_{870}\gtrsim1$\,mJy, which are discussed in Stach et al.\ (in preparation). 

To search for emission lines, we use the dirty data cubes, which were constructed at raw  spectral resolution ($\sim13.5$\,km\,s$^{-1}$) and tapered to $0.5''$ FWHM resolution to match the continuum maps by applying a $\sim400$\,k$\lambda$ Gaussian \emph{uv} taper. These cubes were first continuum-subtracted by subtracting a linear fit to the continuum in the spectrum of each pixel\footnote{Second order polynomial and constant fits were also tested but the linear fit produced a good fit to the data without over-fitting.}. These continuum-subtracted cubes were also used to calculate {\CII} emission sizes in Section \ref{sec:sizes}. 

To search for emission lines in the data cubes, we velocity-binned the $0.5''$ tapered, continuum-subtracted cubes to 50\,km\,s$^{-1}$, 100\,km\,s$^{-1}$, and 200\,km\,s$^{-1}$ channels and then extracted the spectra at the position of each AS2UDS continuum-selected SMG. We search each of these spectra for peaks with ${\rm S/N} \ge2$. These were then refit in the 50\,km\,s$^{-1}$ channel spectrum using a Gaussian profile and the integrated S/N calculated within the FWHM of the line. 

Given the non-Gaussian nature of the noise in the ALMA data cubes, to determine the purity of the sample and hence an acceptable S/N threshold for our detections, we calculate an empirical false positive rate by applying the same procedure to the inverted, velocity-binned, continuum-subtracted cubes at the AS2UDS source positions. This false positive rate is dependent on the velocity binning of the spectra. We require a threshold for selection that produces a false detection rate of ten percent. 

For a false positive rate of ten percent in our final sample, we take an integrated S/N cut which varies depending upon the velocity binning with ${\rm S/N}=8.0$ for $50$\,km\,s$^{-1}$ channels, ${\rm S/N}=7.5$ for $100$\,km\,s$^{-1}$ channels and ${\rm S/N}=7.0$ for $200$\,km\,s$^{-1}$ channels. We detect significant emission lines in ten SMGs above these limits. We plot all ten line emitters in Figure \ref{fig:spectra} and list their properties in Tables \ref{table:lineprops} and \ref{table:photometry}. 

To ensure we identify all bright line emitters within the ALMA pointings, we run two additional line searches. 
First, two of the SMGs where we have identified an emission line lie in ALMA maps which contain a second SMG. In these cases we extract the spectra of this second SMG and search for (lower-significance) emission lines at similar frequencies to the detected emission line. In one of these SMGs (AS2UDS.0109.1) we found a tentative ${\rm S/N}=5.3$ emission line corresponding to the same redshift as the detected emission line source (AS2UDS.0109.0). We include this SMG as a ``supplementary'' source\footnote{The false positive rate at this significance and line width is $\sim50\%$ and so this source requires further observations in order to confirm it.}  and the spectra and optical/near-infrared thumbnails are shown in Appendix \ref{app:UDS109}. 
If confirmed, this secondary source would be located 70\,kpc in projection and 50\,km\,s$^{-1}$ offset in redshift from the primary source. 

Second, we also searched the $695$ ALMA cubes for emission line sources lacking continuum counterparts.  
In each continuum-subtracted cube we step through the cube, collapsing in velocity bins of $100$\,km\,s$^{-1}$ ($\sim7$ resolution elements), the centres of which are shifted by $50$\,km\,s$^{-1}$ between slices\footnote{We also tested channels of $50$ and $200$\,km\,s$^{-1}$ with step size half the channel size, but no additional significant emission lines were detected.}. In each collapsed $100$\,km\,s$^{-1}$ slice we search in the narrow-band image for peaks above $2\sigma$ within the ALMA primary beam. For any peak detected we extract and fit a Gaussian profile to the full spectrum and measure the integrated S/N of the line. We perform the same search on the inverted cubes to calculate a false positive rate. Using this method we find no additional emission line sources within the ALMA pointings above a false positive rate of 50\,percent, corresponding to a line flux limit of $Sdv\gtrsim$1\,Jy\,km\,s$^{-1}$.

\subsection{Multi-wavelength data}\label{sec:multiwavelengthdata}
The UDS has photometric coverage spanning the optical, {near-}, {mid-} and far-infrared, out to radio wavelengths. 

The UKIRT Infrared Deep Sky Survey \citep[UKIDSS;][]{Lawrence2007} UDS data release $11$ (DR11) photometric catalogue (Almaini et al.,\ in preparation) is based on a deep, $K$-band selected catalogue down to a 3\,$\sigma$ depth of $K=25.9$\,mag, with additional imaging in $U$, $B$, $V$, $R$, $I$, $J$, $H$, $K$, and \emph{Spitzer}/IRAC. 

To derive the photometric properties of our sample, we match our sources to the UDS DR11 using a search radius of $0.6$\,arcsec, giving a false-match rate of $3.5$ percent (An et al., submitted). 
Three-colour $5''\times5''$ thumbnails of the ten candidate line emitters are shown in Figure \ref{fig:thumbs}. This figure also shows a zoomed $3''\times3''$ optical/infrared image of each source, with the ALMA continuum contours overlaid. 
We use these thumbnails to assess the multi-wavelength properties of the line emitters, in particular to determine if the optical/infrared photometry is contaminated by nearby galaxies. 

The UDS20 project (Arumugam et al.,\ in preparation) imaged the UDS at $1.4$\,GHz using the Very Large Array. The total area coverage is $\sim1.3$\,degrees$^2$, with the $\sim160$\,hour integration resulting in an rms noise of $\sim6\mu$Jy\,beam$^{-1}$ across the full field. A full description of the radio data will be presented in Arumugam et al. (in preparation). 

The UDS also has coverage with the \emph{Herschel Space Observatory} Photoconductor Array Camera and Spectrometer \citep[PACS;][]{Poglitsch2010} and Spectral and Photometric Imaging REceiver \citep[SPIRE;][]{Griffin2010} at $100$\,\micron, $160$\,\micron, $250$\,\micron, $350$\,\micron, and $500$\,\micron. The resolution of the far-infrared \emph{Herschel} wavebands ($15-35$\,arcsec) requires the data to be deblended in order to obtain the photometry of our SMGs. 

For the deblending we follow the method described in \citet{Swinbank2014}.\footnote{The deblended catalogs for the fields are available from \href{http://astro.dur.ac.uk/~ams/HSOdeblend}{http://astro.dur.ac.uk/$\sim$ams/HSOdeblend}.} 
The deblending uses a combination of the ALMA-detected SMGs and {\it Spitzer}/MIPS 24\,\micron\ and UDS20 radio sources as positional priors for the deblending of the low resolution SPIRE maps.  
To deblend the SPIRE maps we use a Monte Carlo algorithm which fits the observed flux distribution with beam-sized components at the position of each source in the prior catalog. This is then iterated towards solutions that yield the range of possible fluxes associated with each source. To ensure that we do not ``over deblend," the method is first applied at 250\,\micron. Any sources in the prior catalog that are detected at $250$\,\micron\ above $2\sigma$ are then used as the prior list for the 350\,\micron\ deblending, and similarly those detected above $>2\sigma$ at $350$\,\micron\ are then used in the 500\,\micron\ deblending. 
There are an average of $2.4$, $2.0$ and $1.9$ priors within the FWHM of the beam centred at the ALMA position (i.e., $15$\,arcsec, $25$\,arcsec and $35$\,arcsec at $250$\,\micron, $350$\,\micron, and $500$\,\micron\ respectively). 
By attempting to recover false positives injected into the maps we derive $3\sigma$ detection limits of $7.0$, $8.0$, and $10.6$\,mJy at $250$, $350$, and $500$\,\micron, respectively (see \citealp{Swinbank2014} for details). The ALMA sources are included at all wavelengths so as not to bias their SEDs. We discuss the \emph{Herschel} fluxes and far-infrared SED fits in more detail in Section \ref{sec:ciideficit}. 

\begin{table*}
	\centering
	\caption{Table of emission line candidates and line properties. }
	\label{table:lineprops}
	\begin{tabular}{lcccccccc} 
		\hline
		\noalign{\vskip0.05cm}
		Source ID\footnotemark[1] & R.A. & Dec. & $S_{870}$\footnotemark[2] &  $\nu_{\rm obs}$\footnotemark[3] & $z_{\text{\Cii}}$\footnotemark[4]  & FWHM$_{\text{\Cii}}$\footnotemark[5]   &   $S\,dv_{\text{\Cii}}$ & S/N\footnotemark[6]    \\
	& \multicolumn{2}{c}{(J2000)} &      (mJy)      &   (GHz)  &         & (km\,s$^{-1}$) &  (Jy\,km\,s$^{-1}$)     &           \\
		\noalign{\vskip0.05cm}
		\hline 
AS2UDS.0002.1 & 02:18:24.24 & $-$05:22:56.9 & 7.4$\pm$0.5 & 338.707 & 4.611$\pm$0.009 & 220$\pm$50  &   1.3$\pm$0.3  &  8.7 \\      
AS2UDS.0051.0 & 02:19:24.84 & $-$05:09:20.8 & 6.3$\pm$0.4 & 350.571 & 4.421$\pm$0.006 & 770$\pm$80  &   4.0$\pm$0.4  & 10.5 \\         
AS2UDS.0104.0 & 02:16:22.73 & $-$05:24:53.3 & 5.6$\pm$0.3 & 350.447 & 4.423$\pm$0.007 & 530$\pm$60  &   4.9$\pm$0.6  & 17.3 \\      
AS2UDS.0109.0 & 02:16:18.37 & $-$05:22:20.1 & 5.5$\pm$0.7 & 348.715 & 4.450$\pm$0.007 & 440$\pm$40  &   4.5$\pm$0.4  & 11.3 \\       
AS2UDS.0208.0 & 02:19:02.88 & $-$04:59:41.5 & 4.0$\pm$0.7 & 338.445 & 4.615$\pm$0.009 & 290$\pm$40  &   2.2$\pm$0.3  &  8.1 \\       
AS2UDS.0232.0 & 02:15:54.66 & $-$04:57:25.6 & 4.6$\pm$0.3 & 349.140 & 4.443$\pm$0.008 & 340$\pm$90  &   0.9$\pm$0.2  &  7.2 \\       
\emph{AS2UDS.0243.0}\footnotemark[7]  & 02:16:17.91 & $-$05:07:18.9 & 4.3$\pm$0.3 & 350.939 & \ldots & 560$\pm$70  &   3.2$\pm$0.4  & 15.5 \\      
\emph{AS2UDS.0535.0}\footnotemark[8] & 02:18:13.30 & $-$05:30:29.1 & 2.4$\pm$0.5 & 339.301 & \ldots & 310$\pm$20  &   3.6$\pm$0.2  & 12.1 \\         
AS2UDS.0568.0 & 02:18:40.02 & $-$05:20:05.6 & 1.2$\pm$0.3 & 351.701 & 4.404$\pm$0.009 & 340$\pm$60  &   2.8$\pm$0.5  & 10.3 \\         
AS2UDS.0643.0 & 02:16:51.31 & $-$05:15:37.2 & 2.2$\pm$0.4 & 338.512 & 4.614$\pm$0.007 & 390$\pm$110 &   1.5$\pm$0.4  &  7.2 \\         
        \hline 
                \noalign{\smallskip}                                                                                                                                           
Median values\footnotemark[9] & \ldots & \ldots & 4.4$\pm$0.6   & \ldots & 4.45$\pm$0.03 & 370$\pm$50 & 3.0$\pm$0.1 & 10.4$\pm$1.1 \\               
        \hline
        \noalign{\smallskip}
        \hline 
                \noalign{\smallskip}
         \multicolumn{2}{l}{Supplementary catalogue\footnotemark[10]} &&&&&&&\\
        \noalign{\smallskip}
AS2UDS.0109.1 & 02:16:19.04 & $-$05:22:23.2 & 2.6$\pm$0.6 & 348.653 & 4.45$\pm$0.01 & 270$\pm$40 & 1.6$\pm$0.2 &  5.3 \\        \hline     
	\end{tabular}
	
\footnotetext[1]{Source IDs, coordinates and $870$\,\micron\ flux densities come from the full AS2UDS catalogue presented in Stach et al.\ (in preparation).}
\footnotetext[2]{The continuum flux densities are primary beam-corrected and were measured in $1$\,arcsec diameter apertures in the $0.5$\,arcsec FWHM tapered maps.}
\footnotetext[3]{Observed frequencies correspond to the peak of the detected emission line.}
\footnotetext[4]{Redshifts are derived assuming the detected emission line is {\Cii}.} 
\footnotetext[5]{The FWHM and flux density (and their respective uncertainties) of each line are measured from a Gaussian fit to the emission line.}
\footnotetext[6]{S/N measurements come from integrating the spectrum across the line between $\nu_{\rm obs} - 0.5\times{\rm FWHM}$ and $\nu_{\rm obs}+0.5\times{\rm FWHM}$.}
\footnotetext[7]{AS2UDS.0243.0 has optical, near-infrared and radio properties which may indicate the line we detect is CO(8--7), corresponding to $z_{\rm CO} = 1.63\pm 0.01$.}
\footnotetext[8]{AS2UDS.0535.0 has optical and near-infrared properties which may indicate the line we detect is CO(5--4), corresponding to $z_{\rm CO} = 0.70\pm 0.01$.}
\footnotetext[9]{Uncertainties on median values are the standard error.}
\footnotetext[10]{Supplementary sources are those within the same ALMA map as a detected line emitter (but not detected above our S/N threshold), which appear to have a low-significance emission line at a similar frequency to their detected companion.}

\end{table*}

\begin{table*}
	\centering
	\caption{Photometric properties of line emitters. 
	}
	\label{table:photometry}
	\begin{tabular}{lccccccccc} 
	\hline
	\noalign{\vskip0.05cm}
	Source ID     &   $V$\footnotemark[1]    &   $K$    &   $4.5$\micron\  &$S_{250}$&$S_{350}$&$S_{500}$& $S_{1.4 {\rm GHz}}$  &   $z_{\text{phot}}$   & Potential \\ 
	                     & (mag)   &   (mag)  &  (mag)                & (mJy)      &  (mJy)      &       (mJy)   & ($\mu$Jy)            &      &   contamination?\footnotemark[2]  \\ 
	\noalign{\vskip0.05cm}
	\hline
\noalign{\smallskip}
{\it AS2UDS.0002.1} & {\it 26.85$\pm$0.25}  & {\it 23.97$\pm$0.06} 	& {\it 22.76$\pm$0.02}  &  {\it 31$\pm$4} & {\it 35$\pm$5} & {\it 43$\pm$7}   &  {\it $<$80} 	  & \ldots                       & Y: lens? \\ 
AS2UDS.0051.0   	& $>$27.47       	& 23.32$\pm$0.04 		& 21.93$\pm$0.02 		&  $<$9 & $<$11 & $<$12                     &     $<$80 		  &  \ldots 					 & N  \\ 
AS2UDS.0104.0  	    & \ldots  		    	& 24.03$\pm$0.06 		& 23.50$\pm$0.08		&  $<$9 & $<$11 & $<$12                     &       $<$80 	  &  \ldots	                     & N  \\  
AS2UDS.0109.0   	& \ldots  		       	& 22.88$\pm$0.03 		& 22.56$\pm$0.07		&  $<$9 & 11$\pm$3 & $<$14               &       $<$80 	  & \ldots		                 & N \\ 
{\it AS2UDS.0208.0} & {\it 26.11$\pm$0.10}  & {\it 22.94$\pm$0.01}	& {\it 21.28$\pm$0.01}  &  {\it $<$9} & {\it $<$12} & {\it $<$12}                     &  {\it $<$80} 	  & \ldots	                     & Y: lens? \\  
AS2UDS.0232.0  	    &   \ldots  	       	&   \ldots  			&   22.42$\pm$0.01  	&  24$\pm$4 & 20$\pm$4 & $<$12         &       $<$80 	  &  \ldots    & N\\ 
{\it AS2UDS.0243.0} & {\it 23.21$\pm$0.01}  & {\it 20.68$\pm$0.01}  & {\it 20.17$\pm$0.01}  &  {\it 32$\pm$5} & {\it $<$17} & {\it $<$17}               &{\it 1220$\pm$30}& {\it 1.58$_{-0.05}^{+0.05}$} & Y: low-$z$ CO?  \\ 
{\it AS2UDS.0535.0} & {\it 25.95$\pm$0.09}  & {\it 23.44$\pm$0.04} 	& {\it 22.98$\pm$0.07}  &  {\it 12$\pm$3} & {\it 13$\pm$3} & {\it $<$15 }        &     {\it $<$80} & {\it 0.80$_{-0.03}^{+0.03}$}\footnotemark[3] & Y: low-$z$ CO? \\ 
AS2UDS.0568.0  	    & $>27.8$  	        	& 24.36$\pm$0.06 		& 23.39$\pm$0.05 		&  $<$18 & $<$16 & $<$12                     &       $<$80 	  & 3.5$\pm$1.0 	     & N \\  
{\it AS2UDS.0643.0} & {\it $>27.8$} 		& {\it 24.30$\pm$0.03} 	& {\it 21.80$\pm$0.01}  &  {\it 14$\pm$3} & {\it 11$\pm$3} & {\it $<$13}         & {\it 105$\pm$18}& {\it 4.4$_{-1.1}^{+0.6}$} & Y: lens?  \\   

        \hline 
        \noalign{\smallskip}
Median values\footnotemark[4]       & 26.11$\pm$0.33        & 23.44$\pm$0.13        & 22.49$\pm$0.10       & 24$\pm$2  &  13$\pm$2 &  $<13$        & $<80$          & \ldots                       & \ldots \\
        \hline
        \noalign{\smallskip}
        \hline 
        \noalign{\smallskip}
        \multicolumn{2}{l}{Supplementary catalogue} &&&&&&&&\\
        \noalign{\smallskip}

AS2UDS.0109.1       & \ldots                & 24.02$\pm$0.07        & 23.28$\pm$0.13      &  $<$9 & $<$11 & $<$12    &     $<$80       &  \ldots                      & N \\ 

\hline
	\end{tabular}           
	
\footnotetext[1]{Photometry and redshifts are taken from the UDS DR11 catalogue (Almaini et al.,\ in preparation) and the UDS20 radio catalogue (Arumugam et al.,\ in preparation). Ellipses indicate no photometric coverage.}
\footnotetext[2]{Italics indicates that the SMG has a nearby source that may contaminate the photometry. Some or all of these sources may also be lensed (final column); see Section \ref{sec:COvCII}.}
\footnotetext[3]{AS2UDS.0535.0 has a secondary peak in its photometric redshift distribution at $z = 4.63$ (see Appendix \ref{app:notes}).}
\footnotetext[4]{Uncertainties on median values are the standard errors.}
                                                           
\end{table*}

\section{Results and discussion}\label{sec:results}
We identify emission lines in ten AS2UDS continuum sources: three with integrated ${\rm S/N}>7.0$ in the $200$\,km\,s$^{-1}$ channel spectra, six with integrated ${\rm S/N}>7.5$ at $100$\,km\,s$^{-1}$, and one with integrated ${\rm S/N}>8.0$ at $50$\,km\,s$^{-1}$. Figure \ref{fig:spectra} shows the spectra of these sources binned to $100$\,km\,s$^{-1}$ channels (we also show the data at the native resolution). We provide the source redshifts and line properties in Table \ref{table:lineprops}.  
The line flux densities are calculated from the Gaussian profile fit to each line. 

The number of line emitters we identify from the parent sample of $695$ SMGs is consistent with the expectation from the ALESS survey, where two emission line sources were identified from a sample of $99$ SMGs \citep{Swinbank2012}.

\subsection{Alternative emission lines} \label{sec:COvCII}
Before we discuss the properties of our line-emitter galaxies, we first discuss the identification of the emission lines. 
Within the ISM of dusty star-forming galaxies, the brightest emission line in the rest-frame far-infrared is expected to be {\CII}\,$\lambda157$\,\micron. At observed frame $870$\,\micron\ this would correspond to $z\sim4.5$. 
{\CII} dominates the cooling of the ISM for temperatures $T<100$\,K and, as noted earlier, may contribute up to two percent of the bolometric luminosity \citep[e.g.,][]{Smail2011}. 
However, there may be contamination from other emission lines in our sample such as {[N\,{\small II}]}\,$\lambda$122\,\micron\ at $z\sim6.1$, {[O\,{\small I}]}\,$\lambda$145\,\micron\ at $z\sim4.9$, {[N\,{\small II}]}\,$\lambda$205\,\micron\ at $z\sim3.1$ or high-\emph{J}$^{12}$CO at $z=0.3$--$2.7$ ($4<J_{\rm up}<11$). In typical sources the {\CII} emission line is expected to be $\gtrsim10$ times brighter than these other lines \citep[e.g.,][]{Brauher2008} and so we expect contamination to be modest given the shallow depth of the current ALMA data. 

We investigate potential contamination using the multi-wavelength data available in the UDS field. 
The photometric properties of our emission-line SMGs are given in Table \ref{table:photometry}. Most sources are very red or undetected in the optical/near-infrared, which is consistent with them being $z>4$ dusty galaxies. In addition, only two have detections at $1.4$\,GHz, again consistent with the majority being at $z\gg3$ \citep{Chapman2005}. 
A discussion of each of the individual line emitters is given in Appendix \ref{app:notes}.  

Galaxies at $z\sim4.5$ are not expected to be detected in the optical $B$-band due to the Lyman limit at $912$\,\AA\ redshifting to $\gtrsim5000$\,\AA. 
In Figure \ref{fig:thumbs} we show the high-resolution ALMA 870\,\micron\ continuum emission contoured over a $B$-band (or $K$-band) image of each galaxy. Half of the ALMA detections do not have a $B$-band counterpart and/or have photometric redshifts consistent with a $z>4$ galaxy. 
The other five line emitters have $B$-band counterparts which are offset by $\lesssim1$\,arcsec from the ALMA continuum emission. In these cases, we have flagged the photometry and note that this may indicate lensing of the submillimeter source by a foreground galaxy. These five sources are listed in italics in Table \ref{table:photometry} and by circle symbols in all figures where the sources are individually plotted. 

On the basis of their multi-wavelength properties, three of these five sources with nearby $B$-band counterparts appear to be potentially lensed high-redshift {\CII} emitters, as the $B$-band emission is not spatially coincident with the submillimeter emission. We crudely estimate that the lensing of these sources may affect our measured fluxes by a factor of $\lesssim1.5$--$2$, however with the current data we are unable to estimate more precise magnification factors.

\begin{figure}
\includegraphics[width=\columnwidth]{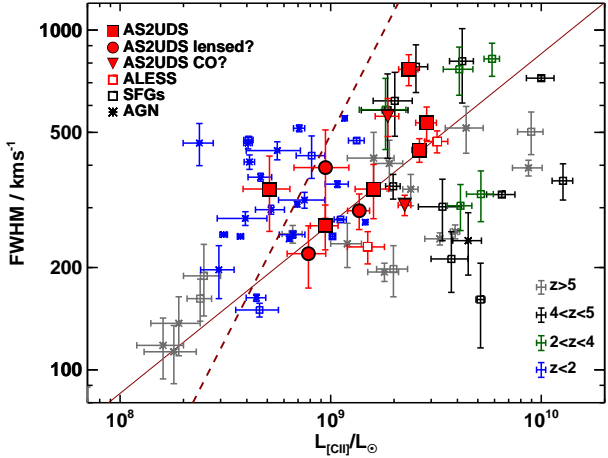}
\caption{{\Cii} FWHM versus line luminosity for the AS2UDS {\Cii} emitters. We compare our sample to star-forming galaxies and AGN from the literature. 
The dashed line shows the $1\sigma$ limit for false positives in the AS2UDS sample (i.e.\ $84$\,percent of false positives lie to the left of this line). 
The solid line shows a ${\rm FWHM}\propto L_{\rm [CII]}^{0.5}$ relation scaled to the median of the AS2UDS data points. This roughly reproduces the trend between the dynamics and luminosity seen in the data, which suggests the line luminosity reflects the mass of the system. 
Note that potentially lensed sources do not appear offset in luminosity at fixed FWHM, suggesting any lens amplification is modest.  
The literature values come from \citet{Iono2006,Venemans2012,Wang2013,Farrah2013,DeBreuck2014,Gullberg2015,Willott2015b,Miller2016,Venemans2016}. 
}
\label{fig:fwhmlcii}
\end{figure}

We next estimate the line luminosities of the two sources where the submillimeter emission is spatially coincident (within $0.5''$) with a $B$-band detection: AS2UDS.0243.0 and AS2UDS.0535.0, assuming these correspond to high-$J$\,$^{12}$CO lines. We compare these luminosities to other studies of high-$J$\,$^{12}$CO emission lines to see whether it is plausible that these lines are high-$J$\,$^{12}$CO rather than {\CII}. 

The photometric redshifts of AS2UDS.0243.0 and AS2UDS.0535.0 are reported in Table \ref{table:photometry}. 
At these redshifts the emission lines would correspond to CO\,($8-7$) at $z=1.63\pm0.01$ with  $L_{{\rm CO}(8-7)} = 1.5\times10^8$\,\Lsun\ for AS2UDS.0243.0 and CO\,($5-4$) at $z=0.70\pm0.01$ with $L_{{\rm CO}(5-4)}  = 0.2\times10^8$\,\Lsun\ for AS2UDS.0535.0. 
These luminosities are approximately an order of magnitude brighter than found in typical local ULIRGs \citep[e.g.\ Arp\,220;][]{Rangwala2011} or AGN-dominated sources \citep[e.g.\ Mrk\,231;][]{vanderWerf2010}. However, recent studies have found comparably luminous sources at higher redshifts \citep[$z>2$, e.g.][]{Barro2017,Yang2017}. 
It is therefore possible that these two sources lie at $z<4$, although they require further investigation to confirm the identity of the emission lines. 

The majority ($\gtrsim80$ percent) of our detected emission lines appear to be {\CII} at $z\sim4.5$. Two have multi-wavelength properties that suggest they are lower-redshift CO emission. However, with no additional detected lines we cannot confirm the identity of the line emission in any of these sources. We therefore proceed with our analysis using all ten line emitters. To guide the reader, in all plots we highlight SMGs which may lie at $z<4$ or whose photometry could be affected by lensing. We also list these sources in italics in all tables. We have tested our conclusions by removing the two potential CO emitters and incorporate the removal of these sources into our error estimates. We find that the majority of our conclusions do not qualitatively change and our estimated quantities do not vary by more than the quoted errors. Any differences are noted in the relevant sections. 
%

\begin{figure}
\includegraphics[width=\columnwidth]{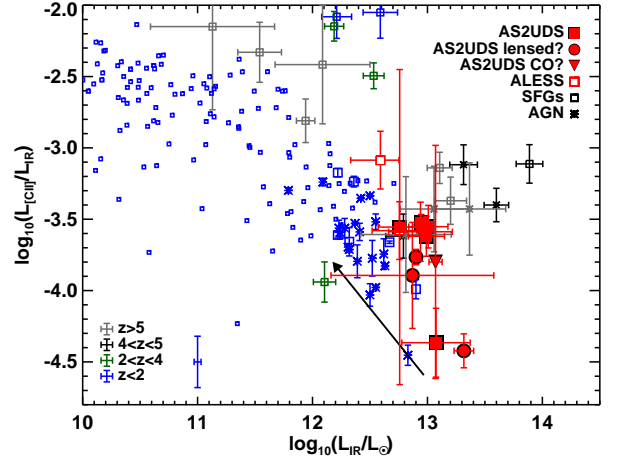}
\caption{Ratio of {\Cii} luminosity to total IR luminosity assuming a fixed dust temperature of $50$\,K for the AS2UDS {\Cii} emitters compared to samples of AGN and star-forming galaxies (SFGs) at $0<z<6.4$ from the literature. Also plotted are the ALESS sources from \citet{Swinbank2012} \citep[with updated luminosities from][]{Swinbank2014}. 
Our new $z\sim4.5$ sample display a continuation of the local trend of decreasing {\Cii} contribution to the total infra-red luminosity towards higher luminosities, contrary to previous high redshift studies. 
The arrow shows the effect of a change in dust temperature from $50$\,K, the median of the AS2UDS sample, to $35$\,K, commonly assumed in $z=2$ SMG studies. 
The literature sample come from \citet{Farrah2013,Brisbin2015,Gullberg2015,Capak2015}. Low redshift ($z<2$) sources shown by small symbols are taken from \citet{Gullberg2015}. A typical uncertainty for these low-redshift sources is shown in the lower left.
}
\label{fig:ciideficit}
\end{figure}

\subsection{{\rm \Cii} luminosities and line widths}
Figure \ref{fig:fwhmlcii} shows the FWHM and emission line luminosities of our ten sources (with those that are potentially lower redshift or lensed flagged) compared to other studies of high redshift star-forming galaxies and AGN. The FWHMs of the lines have a range of $200$--$800$\,km\,s$^{-1}$, with a median of $370\pm50$\,km\,s$^{-1}$. This is similar to the values found in other high redshift studies: \citet{Wang2013} estimate an average FWHM of $\sim360$\,km\,s$^{-1}$ in a sample of five $z>6$ quasars and \citet{Gullberg2015} measure a range of $210$--$820$\,km\,s$^{-1}$ in $20$ strongly-lensed star-forming galaxies at $2.1<z<5.7$. Three of our ten sources have a FWHM of $>500$\,km\,s$^{-1}$, which is a similar fraction to that measured in \citet{Gullberg2015}. 

As discussed in Section \ref{sec:almadata}, the false-positive rate for our sample is a function of velocity binning and line FWHM. In Figure \ref{fig:fwhmlcii} we therefore plot the $1\sigma$ limit for false positive emission lines in our sample (shown by the dashed line). This is determined from the inverted spectra as described in Section \ref{sec:almadata}; $84$\,percent of the false positive emission lines lie to the left of this line. 

The {\CII} luminosity reflects a mixture of the mass of the gas reservoir and the star-formation rate of the SMGs \citep[though see][]{Fahrion2017}, whereas the FWHM is expected to trace their dynamical mass. 
The solid line in Figure \ref{fig:fwhmlcii} shows a model assuming the line luminosity primarily traces dynamic mass: a FWHM$\propto L^{0.5}_{\text{\Cii}}$ relation scaled to the median of the AS2UDS data points. This roughly reproduces the trends seen in the data, albeit with large scatter.

\begin{figure}
\includegraphics[width=\columnwidth]{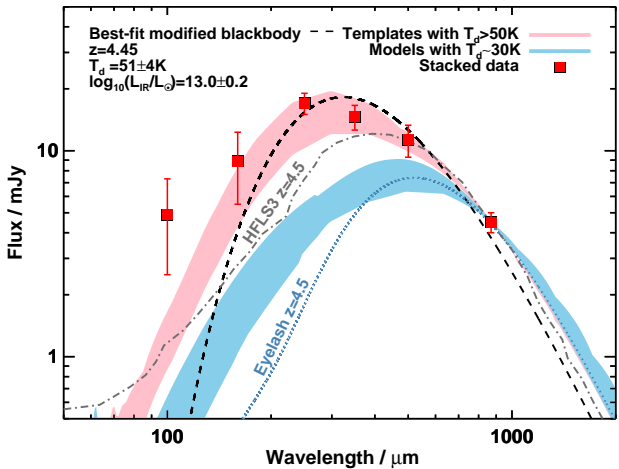}
\caption{The composite infrared SED from \emph{Herschel} PACS/SPIRE and ALMA for our ten line emitters with uncertainties determined from a bootstrap analysis. The dashed black line shows the best modified blackbody fit to this photometry with $\beta=1.8$. The best fit parameters are shown in the top left. Overlaid are shaded regions showing the region occupied by template SEDs with temperatures $T_{\rm d} > 50$\,K and $28$\,K$<T_{\rm d}<35$\,K. The SED of our AS2UDS SMGs at $z\simeq4.5$ suggests warm dust temperatures for these high-redshift galaxies. 
Also plotted are the SEDs of two well-studied SMGs, SMMJ2135$-$0102 \citep[the Eyelash; ][]{Swinbank2010a} and HFLS3 \citep{Riechers2013,Cooray2014}, redshifted to $z=4.5$ and normalized to the average flux density of our sample at $870$\,\micron. }
\label{fig:SED}
\end{figure}

\subsection{{\rm \Cii} deficit} \label{sec:ciideficit}
Previous studies of local star-forming galaxies have found that the ratio of the {\CII} to infrared luminosities declines towards higher infrared luminosities, resulting in a ``{\CII} deficit" in local ULIRGs \citep[Figure \ref{fig:ciideficit}; e.g.,][]{Stacey1991,Malhotra2001,Luhman2003}. Several possible models have been proposed to explain this factor of $\sim100$ decline in {\CII} line luminosity fraction across three orders of magnitude in $L_{\rm IR}$, including enhanced contributions from AGN to the total infrared luminosity in the most luminous galaxies or high-ionization regions contributing more to continuum emission \citep[e.g.,][]{Sargsyan2012,Luhman2003}. At higher redshifts, however, ULIRGs and dusty star-forming galaxies appear to have $L_{\rm \CII}/L_{\rm IR}$ ratios that are comparable to those of less-luminous $z\sim0$ star-forming galaxies \citep[e.g.,][]{Stacey2010,Cox2011,Walter2012,Swinbank2012}. This poses the question of whether the processes responsible for {\CII} emission at $z\gg 0$ differ from those locally. 

To investigate the ``{\CII} deficit" we must first estimate the infrared (rest-frame $8$--$1000$\,\micron) luminosities of our line emitters. To do this we fit a modified blackbody to the far-infrared photometry of each source from \emph{Herschel} PACS, SPIRE and ALMA \citep[e.g., as in][]{Swinbank2014}, adopting a spectral index\footnote{Varying the spectral index over $1.5<\beta<2.3$ changes the luminosities by $\le0.06$\,dex and temperatures by 12\,K (decreasing with larger $\beta$).} of $\beta=1.8$ \citep[e.g.,][]{Bethermin2015}. 

Using this modified blackbody SED we estimate infrared luminosities ranging between $L_{\rm IR} = 7$--$34\times10^{12}$\,{\Lsun} for the AS2UDS emission line sources. The derived characteristic dust temperatures are $39$--$77$\,K. Table \ref{table:luminosity} lists their infrared luminosities, dust temperatures, {\Cii} equivalent widths (EWs) and luminosities. 
The EWs are derived using continuum values from the median of the fit to the full spectrum (derived in Section \ref{sec:almadata}). 

As expected, few of the candidate line emitters are individually detected in the \emph{Herschel} bands (Table \ref{table:photometry}) due to their relative faintness and potentially high redshifts and so, while we can fit the PACS/SPIRE/ALMA fluxes for each individual source with a modified blackbody, these are very uncertain. We therefore also construct an average SED for the whole sample by stacking the individual, un-deblended \emph{Herschel} images at the positions of the ALMA line emitters and extracting bootstrap mean fluxes\footnote{Using bootstrap median fluxes does not change our derived quantities outwith the quoted errors. The infrared luminosities decrease by $0.03$\,dex.}. 
In Figure \ref{fig:SED} we show the stacked sample in the \emph{Herschel} PACS $100$\,\micron, $160$\,\micron\ and SPIRE $250$\,\micron, $350$\,\micron\ and $500$\,\micron\ bands. Fitting a modified blackbody SED to the stack gives a median infrared luminosity $L_{\text{IR}} = (1.0\pm0.4)\times10^{13}$\,\Lsun, assuming the sources lie at $z=4.45$ (the median redshift of our sample), and a median dust temperature of $T_{\rm d} = 51\pm4$\,K, where the uncertainty is taken from a bootstrap analysis which excludes those sources which may be foreground CO emitters. 

Figure \ref{fig:ciideficit} shows the $L_{\rm \CII}/L_{\rm IR}$ ratio versus $L_{\rm IR}$ for the AS2UDS line emitters, compared to local and other high-redshift samples. We note that although our measured {\CII} fluxes (and hence luminosities) have been corrected for the ALMA primary beam, the spectra are measured from the brightest pixel in the tapered map, whereas the infrared luminosities are derived from (aperture-corrected) continuum $1$\,arcsec aperture fluxes (Stach et al.,\ in preparation). As all fluxes are measured from $0.5$\,arcsec tapered images/cubes, the effect of the different methods on our measured line fluxes is likely to be modest. We tested this by extracting spectra from a similar aperture and find, although there is significant scatter between different galaxies, this effect may cause our {\CII} luminosites (and therefore $L_{\rm \CII}/L_{\rm IR}$ ratios) to be low by $0.2$--$0.3$\,dex. 

We compare our sample with local galaxies from \citet{GraciaCarpio2011,Herrera-Camus2018}. These sources include LIRGs and ULIRGs from the Great Observatories All-sky LIRG Survey \citep[GOALS,][]{DiazSantos2013} and normal and Seyfert galaxies from \citet{Brauher2008}. 
The median $L_{\rm \CII}/L_{\rm IR}$ ratio for our $z\sim4.5$ {\CII} emitters $\sim0.02$\,percent, similar to those of local ULIRGs, although the AS2UDS sources have higher infrared luminosities. 
Our measured ratios are in agreement with those of \citet{Swinbank2014} for their two {\CII} emitters at $z=4.4$, however, we measure lower ratios than have been suggested in previous high-redshift studies by \citet{Gullberg2015}, \citet{Capak2015}, or \citet{Brisbin2015}. 
This is a consequence of the relatively warm dust temperatures we estimate for our $z\sim4.5$ SMGs. If we fix the dust temperature for our {\CII} emitters to $\sim35$\,K (as is typically found in $z\sim2$ SMGs) their estimated infrared luminosities decrease by approximately $0.6$\,dex, which is shown by the arrow in Figure \ref{fig:ciideficit}. However, this cooler SED does not match well to the measured photometry (for example at $350$\,\micron\ the required flux density is a factor of two lower than we measure; Figure \ref{fig:SED}, Section \ref{sec:temperatures}). 

The {\CII} deficit is known to be strongly correlated with infrared surface brightness, or equivalently star-formation surface density \citep{DiazSantos2014,Lutz2016,Smith2017}. 
Thus our measured high dust temperatures and large {\CII} deficit may both be explained if our $z\sim4.5$ SMGs have high star-formation rate surface densities, compared to previously-studied sources.

\begin{figure}
\includegraphics[width=\columnwidth]{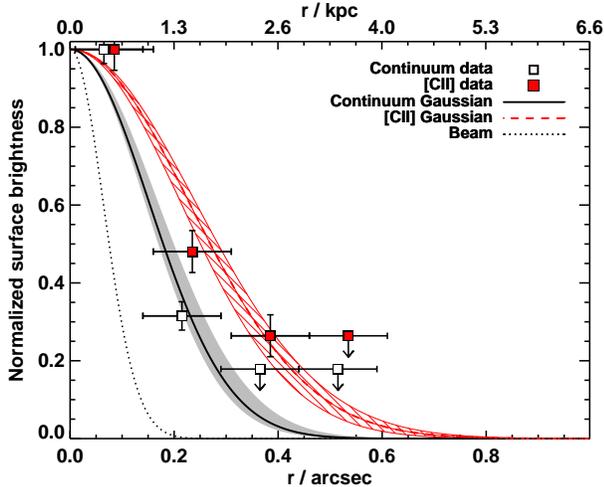}
\caption{Average radial profiles of the dust continuum and {\Cii} emission in our SMGs from the stacked images. The solid black and dashed red lines show the profile derived using a 2D Gaussian fit to the stacked data for continuum and {\Cii} emission respectively. The profiles have been normalized to compare them: the {\Cii} emission is more extended than the continuum emission.  Uncertainties on the Gaussian profiles were determined using a bootstrap analysis. 
Error bars on the abscissae show the size of the radial bins. The points have been offset by $\pm0.01''$ for clarity. $2\sigma$ limits are shown for non-detections. The dotted line shows the ALMA beam profile for reference. 
}
\label{fig:sizes}
\end{figure}

\subsection{Sizes of {\rm \Cii} emitters} \label{sec:sizes}
Previous studies of {\CII} emission line sources have suggested that the {\CII} emission is more extended than the dust continuum emission in high redshift SMGs \citep[e.g.,][with a sample of four $L_{\rm IR}\sim3\times10^{12}$\,\Lsun\ SMGs at $z\sim4.5$]{Gullberg2018}.

{\Nstack} out of our ten line emitters were observed with ALMA in an extended configuration, resulting in maps with synthesised beam FWHM of $\sim0.15''$, or $1$\,kpc at $z=4.5$. We use these data to investigate the sizes of our {\CII} emitters. 
To measure an average size for their dust continuum and {\CII} emission, we stacked the continuum and {\CII} emission of the {\nstack} {\CII} emitters with high-resolution data.\footnote{AS2UDS.0104.0, AS2UDS.0109.0, AS2UDS.0208.0, AS2UDS.0232.0, AS2UDS.0243.0, and AS2UDS.0535.0: the measured profiles do not change if we remove the two potential low redshift emitters.} For the continuum stack, each individual map was normalised to a peak flux of unity to avoid brighter sources dominating the size measurements. {\CII} images were constructed from the emission within $\pm1$\,FWHM in frequency of the emission line peak, continuum subtracted, and then similarly normalised. 
We find no spatial offset between the peaks of the two stacks ($\Delta$dust--{\CII}\,$<0.06$\,arcsec) and no asymmetry in either stack. 
Using a 2-dimensional Gaussian profile fit we measure circularised effective radii (deconvolved with the beam) of $0.15^{+0.02}_{-0.01}$\,arcsec and $0.25^{+0.01}_{-0.04}$\,arcsec  (corresponding to $1.0\pm0.1$\,kpc and $1.7^{+0.1}_{-0.2}$\,kpc at $z\sim4.5$) for the continuum and {\CII} stacks, respectively (Figure \ref{fig:sizes}, where the errors come from bootstrapping). 
This demonstrates that the {\CII} emission is significantly more extended than the continuum emission in these $z=4.5$ galaxies; we find {\CII} sizes similar in extent to the continuum in just $\sim5\%$ of the bootstrap simulations. Our measured sizes are in agreement with those from previous studies at this redshift of SMGs \citep[e.g.,][]{Gullberg2018} as well as quasars \citep[e.g.][]{Kimball2015,DiazSantos2016}, although smaller by a factor of $\sim2$ compared to higher-redshift Lyman break galaxies \citep{Capak2015}. This may indicate that our $z\sim4.5$ SMGs are more compact starbursts than $z>5$ Lyman break galaxies. 

Our sample of ten SMGs has a median infrared luminosity of $(9.3\pm0.4)\times10^{12}$\,\Lsun. This corresponds to a median ${\rm SFR} = 1600\pm200$\,\Msun\,yr$^{-1}$, \citep[using a \citet{Salpeter1955} initial mass function;][]{Kennicutt1998}. With our continuum sizes measured above, this gives a star-formation rate surface density of $130\pm20$\,\Msun\,yr$^{-1}$\,kpc$^{-2}$. This value is slightly higher than that measured in $z\sim2.5$ SMGs \citep[$90\pm30$\,\Msun\,yr$^{-1}$\,kpc$^{-2}$;][]{Simpson2015a} and may indicate that our SMGs follow the trend found in \citet{Smith2017} of increasing \CII\ deficit with star-formation rate surface density. For a review of possible physical explanations we refer the reader to \citet{Smith2017} who find that this is likely caused by local physical processes of interstellar gas rather than global galaxy properties such as total luminosity.

We also fit the sizes of the galaxies in the near-infrared UKIDSS/UDS DR11 $K$-band image, to trace the stellar emission, using GALFIT/GALAPAGOS \citep{Galfit,Galapagos}. Further details of the UDS $K$-band S{\'e}rsic fitting can be found in \citet{Almaini2017} \citep[see also][]{Lani2013}. 
We were able to obtain fits for four of our ten\footnote{AS2UDS.0051.0, AS2UDS.0109.0, AS2UDS.0535.0, AS2UDS.0568.0: the median size does not change if we remove AS2UDS.0535.0.} emitters. The remaining SMGs were either too faint in the $K$-band or were blended with other nearby sources. These four galaxies give a median (deconvolved) size which is considerably more extended than both the dust and {\CII} emission: $0.7\pm0.1$\,arcsec radius ($4.7\pm0.7$\,kpc). This is slightly larger than found in studies of \emph{Hubble Space Telescope} $H_{160}$ sizes of SMGs at $1<z<3.5$, which measure half-light radii of $\sim2.7\pm0.4$\,kpc \citep{Swinbank2010b,Chen2015}. 
The $K$-band fits also output a median S{\'e}rsic index $n=0.80\pm0.06$; consistent with a disk profile. This suggests that the stars visible in the rest-frame UV are more extended than the dust-obscured starburst region in these systems. This may suggest that we are seeing highly dissipative starbursts occurring within pre-existing stellar systems.

\subsection{$z\sim4.5$ SMGs are warm} \label{sec:temperatures}
\begin{figure}
\includegraphics[width=\columnwidth]{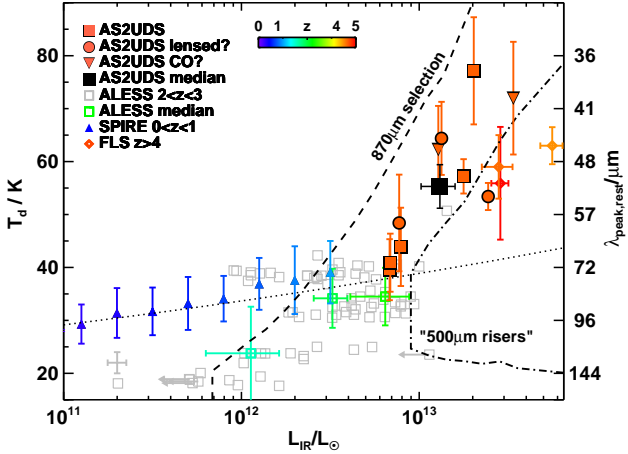}
\caption{Dust temperature ($T_{\rm d}$) as a function of total infrared luminosity ($L_{\rm IR}$) for the line emitters in our survey, colour-coded by redshift. At a given redshift, the characteristic dust temperature roughly corresponds to the wavelength at which a galaxy's SED peaks. Individual line emitters are shown, as well as the median of the sample. The $z\simeq4.5$ AS2UDS candidate {\Cii} emitters have warmer dust temperatures, or equivalently their dust SEDs peak at shorter wavelengths, at fixed luminosity compared to $z\simeq0$ galaxies. 
For comparison we also show ALESS SMGs from \citet{Swinbank2014} (a typical error bar is shown in the lower left), with median values overplotted and colour-coded by redshift. Binned low-redshift SPIRE-selected (U)LIRGs from \citet{Symeonidis2013} are shown by the solid triangles. We indicate the low-redshift correlation by the dotted line. 
The dashed line shows the approximate selection limit for an $870$\,\micron-selected sample with $S_{\rm 870} > 3.6$\,mJy: sources to the right of this line would be detected. The dot-dashed line shows the approximate selection limit for $4<z<6.5$ sources whose SED peaks at $\lambda\ge 500$\,\micron, so-called ``$500$\,\micron\ risers" \citep{Dowell2014}: sources to the right of this line would be selected. Plotted for comparison are three such sources, with spectroscopic redshifts $z>4$, from the First Look Summary (FLS) of the \emph{Herschel} Multi-tiered Extragalactic Survey \citep[HerMES;][]{Oliver2012}. 
}
\label{fig:Td}
\end{figure}

Comparing their inferred dust temperatures and far-infrared luminosities in Figure \ref{fig:Td}, our sample of $z\sim4.5$ SMGs appear to have warmer characteristic dust temperatures at fixed luminosity than inferred for $z\simeq2$ SMGs and star-forming galaxies \citep[e.g.,][but see also \citealt{Casey2012}]{Swinbank2014,Magnelli2012a,Symeonidis2013,daCunha2015}. In this section we first test the reliability of our measured dust temperatures and then discuss the implications of warm dust temperatures on the selection of high redshift SMGs. 

We first note that at $z\sim4.5$ the temperature of the cosmic microwave background (CMB) is $\sim15$\,K and this can bias temperature measurements of high redshift galaxies \citep{daCunha2013}. However, the temperatures we measure for our {\CII} emitters, $T_{\rm d}=39$--$77$\,K, are $2.5$--$5\times$ warmer than the CMB and so this is unlikely to significantly affect our measurements. 

Warmer characteristic dust temperatures correspond to the far-infrared SED peaking at shorter wavelengths. 
For a $z=4.5$ SMG, the \emph{Herschel} PACS/SPIRE wavebands correspond to $\lesssim90$\,\micron\ rest-frame. This means that our measurement of cold dust in these SMGs is dominated by the $870$\,\micron\ flux measurement (rest-frame $160$\,\micron). To test whether having only one datapoint at longer wavelengths may bias our measurements toward warmer measured dust temperatures, we use simulated SEDs for galaxies with a range of known dust temperatures and infrared luminosities, redshifted to $z=4.5$. Convolving these simulated SEDs with the \emph{Herschel}/ALMA observational wavebands and applying the flux limits appropriate for our observations, we then measure ``observed" dust temperatures. We find the measured dust temperatures agree with the input temperatures, with $|T_{\rm d,in}-T_{\rm d,out}|/T_{\rm d,in}=0.07$ on average. Our measured temperatures are therefore not a consequence of the coverage of our observations. 

From Figure \ref{fig:SED} we see that, as with the individual sources, the measured average dust temperature for the composite SED is warm compared to typical temperatures of $\sim32\pm1$\,K estimated for SMGs at lower redshifts \citep{Swinbank2014}. This is due to the higher average infrared luminosity in the $z\sim4.5$ sample; \citet{Swinbank2014} measure $L_{\rm IR} = (3.0\pm0.3)\times10^{12}$\,\Lsun\ at $z\sim2.5$, compared to our value of $L_{\text{IR}} = (1.0\pm0.4)\times10^{13}$\,\Lsun. 
As the far-infrared SED of distant galaxies can be reasonably approximated by modified blackbody emission, the infrared luminosity is proportional to dust mass and temperature as $L_{\rm IR} \propto M_{\rm d} T_{\rm d}^{4+\beta}$. 
We measure similar sizes for the continuum emission in our $z\sim4.5$ SMGs as at $z\sim2.5$: $2.4\pm0.2$\,kpc \citep{Simpson2015a}. 
As infrared luminosity traces star-formation rate, and dust mass is correlated with the radius of the dust-emitting region, the higher average luminosity and similar average radius correspond to a higher star-formation rate surface density, which explains the high temperatures observed in our sample. 

Significant effort has been invested into identifying dusty star-forming galaxies at $z\gtrsim4$ using \emph{Herschel}/SPIRE \citep[e.g.][]{Roseboom2012,Dowell2014,Yuan2015,Asboth2016}. 
We investigate whether our blind {\CII} detections in a $870$\,\micron\ survey would be identified by these other selections. 
Figure \ref{fig:Td} shows two selection limits: 
firstly, SMGs that are brighter than $S_{870}\simeq 3.6$\,mJy (``$870$\,\micron\ selection"). Secondly we show the selection of sources with $S_{500}>30$\,mJy and $S_{500} > S_{350} > S_{250}$, i.e.\ a red ``$500$\,\micron\ riser'' \citep[e.g.,][]{Dowell2014}. 
The latter would select one of our \CII\ emitters (AS2UDS.0002.1, which may be lensed) but would miss most of our sample as they are too faint at $500$\,\micron. If it was possible to go $\sim2$--$3$ times deeper at $500$\,\micron, then the ``$500$\,\micron\ riser'' selection would have identified all our candidate \CII\ emitters.

\begin{table*}
	\centering
	\caption{{\Cii} emitter derived properties.}
	\label{table:luminosity}
	\begin{tabular}{llllll} 
	\hline
	\noalign{\vskip0.05cm}
	Source ID   &   L$_{\text{IR}}$  & L$_{\text{IR,free Td}}$  &   T$_\text{d}$    &   L$_{\text{\Cii}}$   &   EW$_{\text{\Cii}}$\footnotemark[1]   \\
	                   &   ($10^{12}$\Lsun)  &   ($10^{12}$\Lsun)   &      (K)              &    ($10^{8}$\Lsun)      &      (km\,s$^{-1}$)              \\
	\noalign{\vskip0.05cm}
	\hline
{\it AS2UDS.0002.1}\footnotemark[2] &  $21 \pm  4$ & $24\pm 4$             &   $53 \pm  3$  &     8 $\pm$  2   &     200 $\pm$   50  \\
AS2UDS.0051.0        &  $ 10 \pm  3$ & $ 7\pm 3$             &   $40 \pm  6$  &   24 $\pm$  3   &     580 $\pm$   70  \\
AS2UDS.0104.0        &  $10 \pm  3$ & $ 7\pm 3$              &   $41 \pm  6$  &   29 $\pm$  3   &    1050 $\pm$  140  \\
AS2UDS.0109.0        &  $ 9 \pm  1$  & $ 8\pm 1$              &   $44 \pm  7$  &   26 $\pm$  2   &    1100 $\pm$  310  \\
{\it AS2UDS.0208.0}\footnotemark[2]  &  $ 8.0 \pm  0.2$ & $ 7.7\pm 0.2$   &   $48 \pm  9$  &   14 $\pm$  2   &     650 $\pm$  240  \\
AS2UDS.0232.0        &  $12 \pm  6$  & $18\pm 6$             &   $57 \pm  3$  &     5 $\pm$  1   &     210 $\pm$   60  \\
{\it AS2UDS.0243.0}\footnotemark[2]$^,$\footnotemark[3]  &  $<22$  & $34\pm22$                     &   $72 \pm 11$  &  19 $\pm$  2   &     850 $\pm$  160  \\
{\it AS2UDS.0535.0}\footnotemark[2]$^,$\footnotemark[3]  &  $ 9 \pm  4$ & $13\pm 4$             &   $62 \pm  8$  &   23 $\pm$  2   &    2200 $\pm$  800  \\
AS2UDS.0568.0        &  $<6$ & $20\pm15$                       &   $77 \pm 10$  &  16 $\pm$  3   &    2300 $\pm$ 1300  \\
{\it AS2UDS.0643.0}\footnotemark[2] &  $ 7 \pm  6$  & $13\pm 6$              &   $64 \pm  7$  &     9 $\pm$  3   &     700 $\pm$  260  \\
        \hline 
                \noalign{\smallskip}
Median values\footnotemark[4] & $9.3 \pm 0.4$  & 13.1 $\pm$ 0.9 & 55 $\pm$ 4 & 17.3 $\pm$ 0.8 & 770 $\pm$ 70 \\
	\hline
    \noalign{\smallskip}
    \hline 
            \noalign{\smallskip}
     \multicolumn{2}{l}{Supplementary catalogue} &&&&\\
    \noalign{\smallskip}
AS2UDS.0109.1  &  $< 7.0$  &  $< 8.0$   &   $<55$  &    7.6 $\pm$  1.2   &     900 $\pm$  500  \\     
\hline
	\end{tabular}    
	
\raggedright Two values of L$_{\text{IR}}$ are listed: the value obtained from fitting the far-infrared SED with a modified blackbody with a fixed dust temperature of $50$\,K and, in brackets, the value obtained if dust temperature is left as a free parameter. 
\footnotetext[1]{Equivalent widths are given in the  rest frame.}
\footnotetext[2]{Italics indicates that the source has a nearby companion that may contaminate the photometry.} 
\footnotetext[3]{AS2UDS.0243.0 and AS2UDS.0535.0 have photometric properties which indicate the line we detect may instead be CO(8--7) or CO(5--4) respectively, corresponding to $z_{\rm CO} < 2$. Further observations are needed to confirm the nature of these emission lines.}
\footnotetext[4]{Uncertainties on median values are the standard error.}
                                                                  
\end{table*}

\subsection{Molecular and dynamical masses of $z\gtrsim4$ {\rm \Cii} emitters}
SMGs have high star-formation rates and are believed to have large cold gas reservoirs \citep{Bothwell2013}. The mass of cold gas available places constraints on the duration of the starburst, which has implications for the predicted number densities of SMGs. 

The cold gas mass can be estimated from the dust mass of a galaxy by adopting a suitable gas-to-dust ratio. 
Previous studies of the gas-to-dust ratio in star-forming galaxies and SMGs have found a range of values: for example, in local star-forming galaxies the gas-to-dust ratio is higher \citep[$\text{GDR}=130\pm20$;][]{Draine2007}, whereas in the ALESS sample \citet{Swinbank2014} estimate a gas-to-dust ratio of $\text{GDR}=75\pm10$ using the correlation between stellar mass, metallicity and star-formation rate from \citet{Maiolino2008} \citep[see also][]{Magnelli2012b}. 
While, from fitting the far-infrared SEDs of the ALESS sample, \citep{Magnelli2012a} suggest a slightly higher average gas-to-dust ratio of $\text{GDR} = 90\pm25$. 
The gas-to-dust ratio is also a function of metallicity and has been suggested, from studies of a small number of SMGs, to decrease towards higher redshifts \citep[e.g.,][]{Ivison2010d,Santini2010}.

We estimate the dust masses for our SMGs using
\begin{equation}
M_{\rm d} = \frac{S_{870} D_{\rm L}^2}{\kappa B_{\nu}(1+z)},
\end{equation}

\noindent where $B_{\nu}$ is the Planck function, $D_{\rm L}$ is the luminosity distance, $S_{870}$ is the flux density at rest-frame $870$\,\micron, extrapolated assuming $\beta=1.8$, $\kappa = 0.077$\,m$^2$\,kg$^{-1}$  \citep{Dunne2000} and the values of $T_{\rm d}$ and $z_{\rm [CII]}$ derived in Section \ref{sec:ciideficit}.  We obtain an average dust mass $M_{\rm d} = (1.6\pm0.3) \times10^8$\,\Msun, although we note that the measured dust mass is a strong\footnote{The dust mass varies from $0.5$--$2.7\times10^8$\,\Msun\ for $\beta=2.4$--$1.5$.} function of $\beta$ and $T_{\rm d}$. This dust mass is similar to those found in previous surveys at $z\le4$ \citep[e.g.,][]{Swinbank2014,daCunha2015}. 
Adopting the value of $\text{GDR}=90\pm25$ from \citet{Magnelli2012a} then predicts gas masses of $M_{\rm gas} = (1.5\pm0.5)\times 10^{10}$\,\Msun. 

To obtain a gas fraction for our sample, we estimate total masses from dynamical measurements of the {\CII} line. From the median size of the {\CII} emission region in our sample ($r_{\rm e}=1.7\pm0.2$\,kpc), and the median line width of FWHM$=370\pm50$\,km\,s$^{-1}$ (Table \ref{table:lineprops}) we can estimate a dynamical mass using a disk model for the {\CII} line dynamics: 
\begin{equation} \label{eqn:mdyn}
M_{\rm dyn} = \frac{1}{\sin^2(i)} \frac{v^2r_{\rm e}}{G},
\end{equation}

\noindent where $v$ is the line width, $r_{\rm e}$ is the effective radius, and $G$ is the gravitational constant. Assuming the average value for randomly inclined galaxy disks $\langle$sin\,$i \rangle = 0.79$ \citep{Law2009} gives $M_{\rm dyn} = (2.1\pm0.6)\times10^{10}$\,\Msun. 

To scale our measured total gas masses to the gas enclosed within $r_{\rm e}=1.7$\,kpc, and thus directly compare them to the calculated dynamical mass, we assume the cold gas reservoir is similar in scale to the {\CII} emission and so scale the gas mass by the fraction of {\CII} emission within the effective radius: 50\,percent. 
With this we estimate a typical gas fraction of $f_{\rm gas} = 0.5 \times M_{\rm gas} / M_{\rm dyn} = 0.4\pm0.2$. This is consistent with the value predicted for $z=4.5$ galaxies by \citet{Tacconi2018} of $f_{\rm gas} \sim 0.4$ and the $z\sim4$ value estimated in \citet{Bethermin2015} of $f_{\rm gas}=0.65^{+0.16}_{-0.29}$. 
We caution that this value is highly dependent upon the adopted gas-to-dust ratio and measured dust and dynamical masses, each of which have large uncertainties and require further detailed observations to confirm.

\subsection{{\rm \Cii} luminosity function}
The evolution of the number density and luminosity of {\CII} emitters as a function of redshift provides information on the evolution of the star formation rate density. This is particularly useful at $z>4$ where the contribution of dust-obscured galaxies to the cosmic star formation rate density is still relatively unconstrained \citep[e.g.,][]{Madau2014}.

\begin{figure}
\centering
\includegraphics[width=\columnwidth, trim={0cm 0cm 0cm 0cm}, clip]{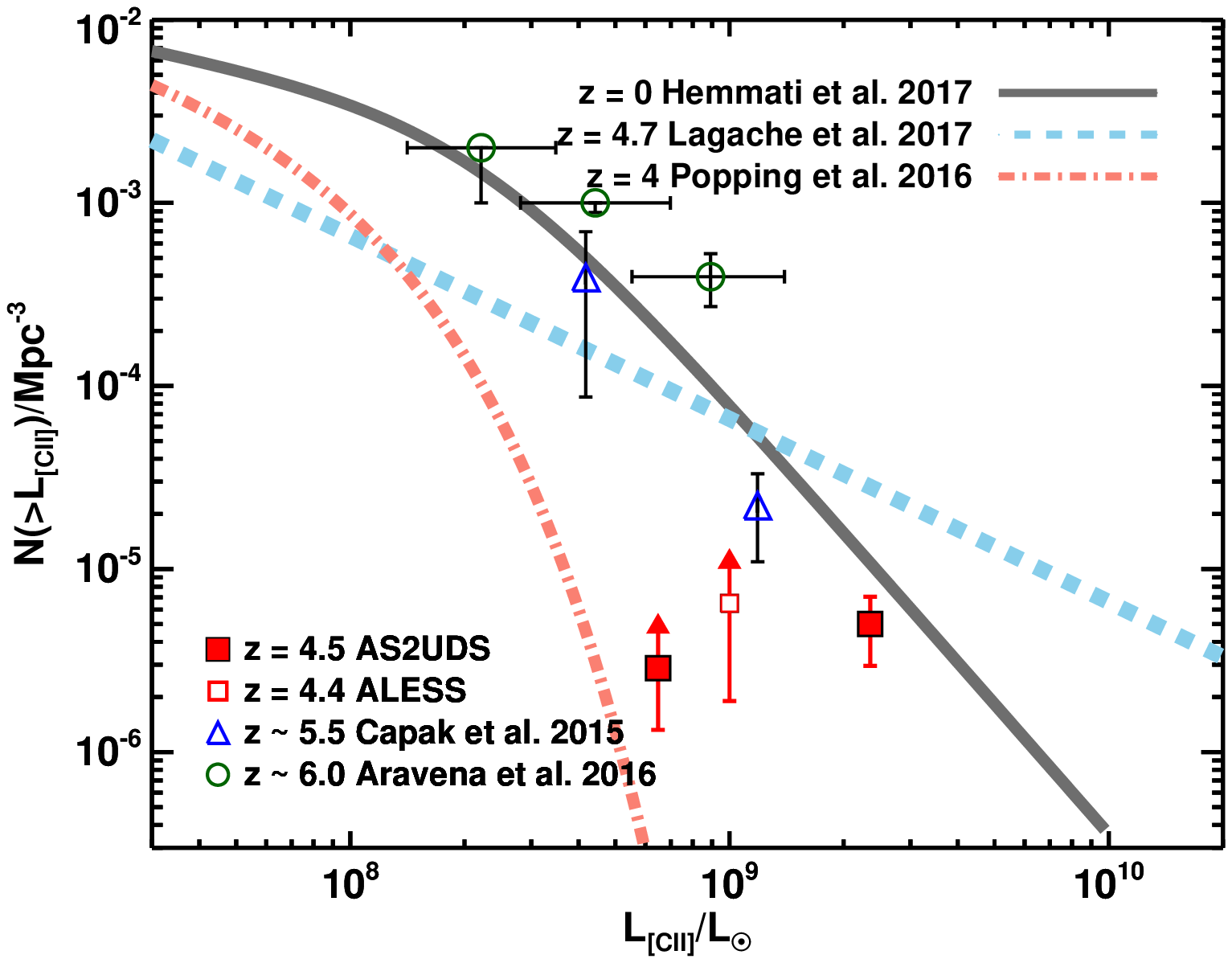}
\vspace{-0.5cm} \noindent
\caption{
The $z\sim4.5$ \Cii\ luminosity function for the AS2UDS continuum-selected \Cii\ emitters compared to the ALESS sample from \citet{Swinbank2012} and higher redshift studies from \citet{Capak2015} and \citet{Aravena2016}. 
Overlaid are the $z=0$ \Cii\ luminosity function measured in \citet{Hemmati2017}, the $z=4$ model from \citet{Popping2016}, and the $z\sim4.7$ model from \citet{Lagache2018}. 
We note that our data points are a lower limit on the \Cii\ luminosity function as our sample are detected in both \Cii\ and the dust continuum. 
Our data are consistent with the measured limit from \citet{Swinbank2012} and with little-to-no evolution in the \Cii\ luminosity function since $z=0$. The \citet{Popping2016} model underpredicts our observations, while the \citet{Lagache2018} model and \citet{Hemmati2017} $z=0$ observations are consistent with our lower limit on the bright end of the \Cii\ luminosity function at this redshift. 
}
\label{fig:ciilfn}
\end{figure}

With our sample of SMGs we can calculate a {\CII} luminosity function at $z\simeq4.5$. Here we only use those with continuum flux densities of $S_{870} \ge 3.6$\,mJy as our continuum survey is complete to this level. We therefore use seven of our SMGs (Table \ref{table:lineprops}). We first correct for incompleteness in our sample: 
from inserting $10^4$ false emission lines into spectra where no lines were originally detected and re-running our emission line detection we determine the recovery rate of emission lines as a function of line luminosity. From this analysis we find completeness rates of $90$\,percent at $L_{\rm [CII]}>2.3\times10^9$\,\Lsun, $80$\,percent at $L_{\rm [CII]}>1.5\times10^9$\,\Lsun, $70$\,percent at $L_{\rm [CII]}>8\times10^8$\,\Lsun\ and $50$\,percent at $L_{\rm [CII]}>3\times10^8$\,\Lsun. Our luminosity function has been corrected for these completeness rates. The lower luminosity bin has a completeness of $\lesssim70$\,percent. We therefore plot a lower limit as our correction is unlikely to account for the full incompleteness. 

We note that our line emitters are not a simple {\CII}-selected sample as they were all selected to have a dust continuum detection. With the current samples of high-redshift SMGs available it is not currently possible to estimate how many \CII\ emitters may be missed due to this selection. Our measured values are therefore likely to be a lower limit on the {\CII} luminosity function. 
Removing the SMGs in our sample that are potential lenses or lower-redshift sources lowers our estimate at the bright end of the luminosity function by $<2\sigma$ and so the following analysis remains qualitatively similar. 

Since our ALMA survey observed all the $S_{\rm 850} \ge 4$\,mJy S2CLS submillimeter detections in the UDS, we use the full area of the field and the number of {\CII} emitters to estimate the bright end of the {\CII} luminosity function at $z\sim4.5$. The UDS is a $\sim0.96$ sq. degree field and our ALMA observations cover a redshift range $\Delta z=0.12$, giving a comoving volume of $1.2\times10^6$\,Mpc$^3$. 
Figure \ref{fig:ciilfn} shows the {\CII} luminosity function at $z\sim4.5$ as derived from our sample of seven $S_{870}>4$\,mJy continuum-selected {\CII} emitters.

Our measured {\CII} luminosity function agrees well with the lower limit derived by \citet{Swinbank2012} and the estimated density of UV-selected $z\sim5.5$ {\CII} emitters from \citet{Capak2015}. We find a lower number density than measured in \citet{Aravena2016}, whose values at $z\sim6$ are upper limits as the emission line sources in their sample are unconfirmed candidate {\CII} emitters.

In Figure \ref{fig:ciilfn} we compare to the $z=0$ {\CII} luminosity function from \citet{Hemmati2017}, the predicted $z=4$ luminosity function from \citet{Popping2016}, and the predicted form at $z=4.7$ from \citet{Lagache2018}. 
We find no evidence for any strong evolution of the {\CII} luminosity function between $z=0$ and $z\sim4.5$, consistent with the prediction from \citet{Hemmati2017} that the {\CII} luminosity function increases from $z=0$ to $z=2$ and decreases again thereafter. 

Our observations at the bright end of the luminosity function suggest that the model of \citet{Lagache2018} overpredicts the {\CII} luminosity function at this redshift unless our sample selection is incomplete, which is possible. By contrast, the \citet{Popping2016} model, as noted in that study, vastly under-predicts the observed number density of {\CII} emitters at all luminosities. This is also true for their models at $z=0$ and $z=6$. 
The under-prediction of the {\CII} luminosity function may be due to the semi-analytic models producing gas reservoirs in galaxies which are much less massive than expected, thus having star-formation rates which are low and therefore lower {\CII} luminosities. 

We can also use our results to place limits on the redshift distribution of SMGs. 
We can use our sample to estimate the number density of $S_{\rm 870}\ge 3.6$\,mJy sources at $z\simeq4.5$ as $6$--$7$ in $1.2\times10^6$\,Mpc$^3$ (where the range indicates removing/retaining AS2UDS.0243.0 in the sample), yielding a space density of $\gtrsim(5$--$6\pm1)\times10^{-6}$\,Mpc$^{-3}$. This indicates that $\sim50$--$60/695$ ($7$--$8\pm4$ percent) of the SMGs in AS2UDS are expected to be at $z=4$--$5$. 
This is consistent with \citet{Simpson2014} who find $35\pm5$\,percent of SMGs in the ALESS survey lie at $z>3$ \citep[see also][ who estimate that $\sim23$\,percent of their spectroscopic SMG sample are at $z>3$]{Danielson2017} and \citet{Michalowski2017} who estimate that $23$\,percent of SCUBA-2 submillimeter sources in the UDS are at $z\ge4$.

\section{Conclusions}\label{sec:conclusions}
We have used ALMA at $870$\,\micron\ to observe a sample of $716$ SCUBA-2-selected sources brighter than $S_{850}\ge3.6$\,mJy ($4\sigma$) in the UKIDSS UDS field. With high resolution ($0.15''$--$0.30''$) ALMA observations we detect the submillimeter galaxy counterparts to the single-dish submillimeter sources. We use these data to identify bright line emitters that fall in the $7.5$\,GHz bandwidth of our ALMA observations. We detect ten line emitters above a signal-to-noise of $7$, corresponding to a false detection rate of ten percent (as derived from inverting the data cubes). All of these line emitters are $870$\,\micron\ continuum-detected sources in our parent survey with $S_{870}\gtrsim1$\,mJy. 
The majority of these line emitters have multi-wavelength properties consistent with {\CII} $\lambda 157.8$\,\micron\ emitters at $z=4.4$--$4.6$. Our main results are summarised below.

\begin{itemize}
\item We detect ten candidate line emitters, at least eight of which are likely to be {\CII} emitters at $z=4.4$--$4.6$, placing a lower limit of $>5\times10^{-6}$\,Mpc$^{-3}$ on the space density of SMGs at $z=4.5$ and suggesting $\ge7$\,percent of $S_{870}\gtrsim1$\,mJy SMGs lie at $z>4$. 
\item The ratio of $L_{\rm {\CII}}/L_{\rm IR}$ for our galaxies is similar to that seen in ULIRGs at $z\simeq0$, even though the SMGs have infrared luminosities an order of magnitude higher. 
\item Through stacking we find that the {\CII} emission is $1.5\times$ more extended than the continuum dust emission in $z\sim4.5$ SMGs (1.7\,kpc versus 1.0\,kpc). The extended {\CII} emission compared to the dust continuum together with the measured infrared luminosities in these SMGs combine to give a star-formation rate surface density measurement of $130\pm20$\,\Msun\,yr$^{-1}$\,kpc$^{-2}$. This high surface density of star formation may explain the large \CII\ deficit, as shown in \citet{Smith2017}, however the origin of the {\CII} deficit is complex and unlikely to be simply caused by one factor.  
\item We find that the {\CII} luminosity function at $z=4.5$ is very similar to that at $z=0$. The similar luminosity functions at $z=4.5$ and $z=0$ are consistent with the prediction from \citet{Hemmati2017} that the \CII\ luminosity function increases up to $z\simeq2$ and decreases again thereafter.  

\medskip
\end{itemize}

\acknowledgements
The authors would like to thank the anonymous referee for their comments which improved the flow and content of this paper. 
EAC, BG and IRS acknowledge support from the ERC Advanced Investigator Grant DUSTYGAL (321334) and STFC (ST/P000541/1). 
IRS also acknowledges support from a Royal Society/Wolfson Merit Award. 
JLW acknowledges an STFC Ernest Rutherford Fellowship. 
MJM acknowledges the support of the National Science Centre, Poland through the POLONEZ grant 2015/19/P/ST9/04010; this project has received funding from the European Union's Horizon 2020 research and
innovation programme under the Marie Sk{\l}odowska-Curie grant agreement No. 665778. 
This paper makes use of the following ALMA data: ADS/JAO.ALMA\#2015.1.01528.S, ADS/JAO.ALMA\#2016.1.00434.S. 
ALMA is a partnership of ESO (representing its member states), NSF (USA) and NINS (Japan), together with NRC (Canada), MOST and ASIAA (Taiwan), and KASI (Republic of Korea), in cooperation with the Republic of Chile. The Joint ALMA Observatory is operated by ESO, AUI/NRAO and NAOJ.

\appendix

\section{Supplementary source AS2UDS.0109.1} \label{app:UDS109}

\begin{figure}[h!] 
\vspace{-0.5cm} \noindent     
\includegraphics[trim= 8.1cm  6.5cm  1.5cm  0.0cm,clip,width=0.18\textwidth]{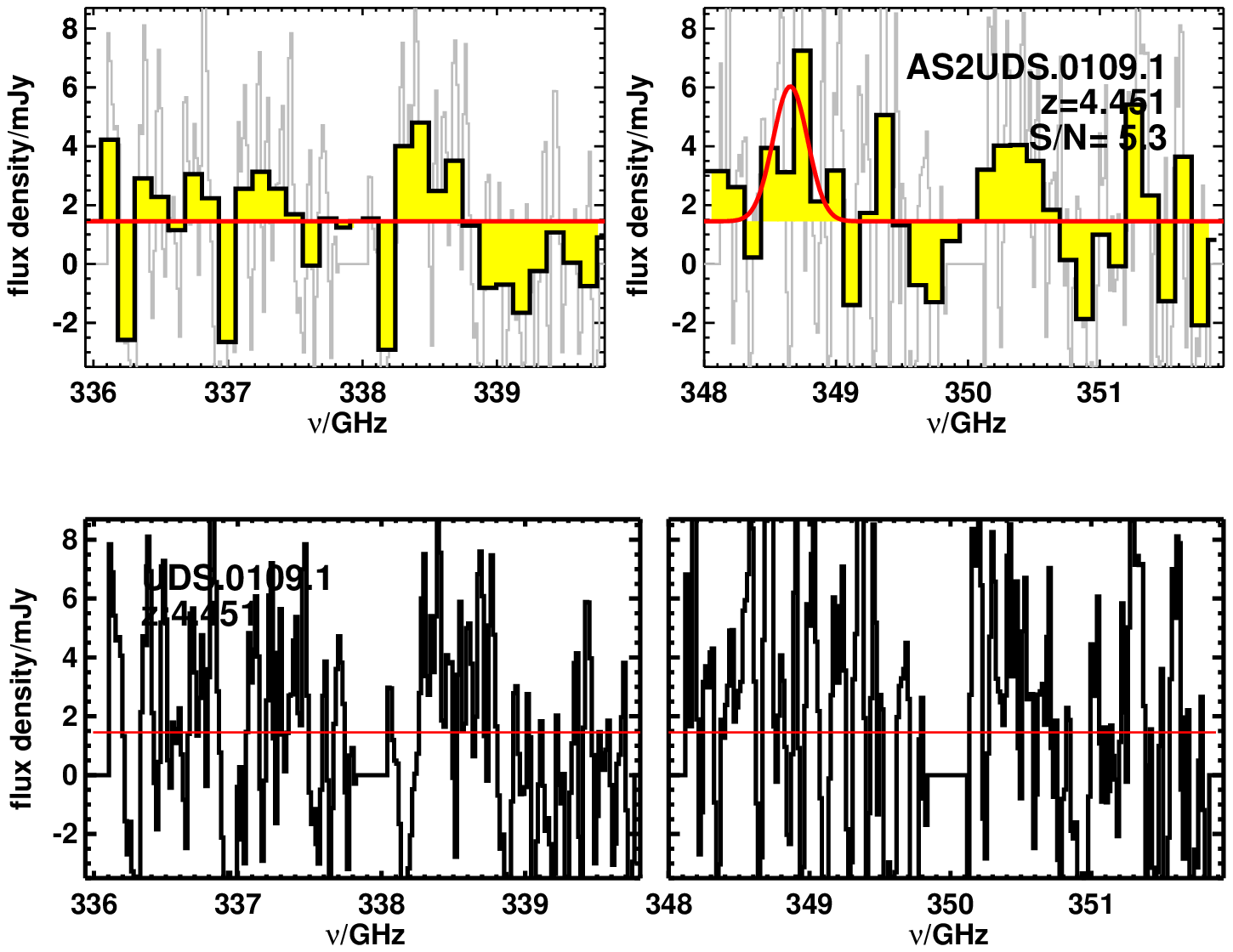}
\includegraphics[width=0.83\textwidth, trim={2.3cm 0.7cm 8cm 1cm}, clip]{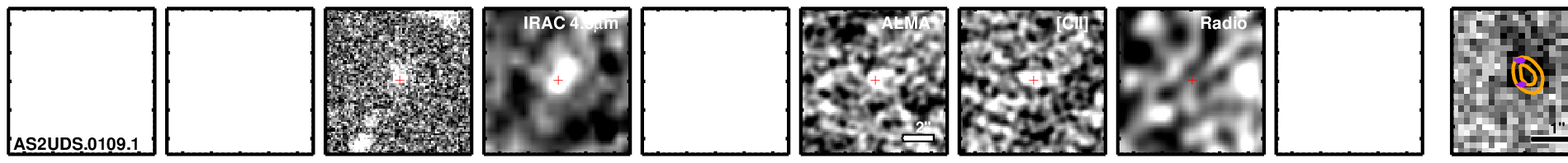}

\caption{
Spectrum and thumbnails for the supplementary \CII\ emitter AS2UDS.0109.1. This is a continuum source in the same ALMA map as AS2UDS.0109.0, an $11\sigma$ line emitter at $z=4.450$. Examining the same region of the spectrum in AS2UDS.0109.1 we find a tentative $5.3\sigma$ detection of an emission line, which would correspond to {\Cii} at $z=4.451$. The photometric thumbnails where available are also shown. Without optical coverage it is difficult to determine, but the $K$ and IRAC photometry is consistent with a $z>4$ source. Further observations are required to confirm this companion source. }
\label{fig:suppl}
\end{figure}

Some of our detected line emitters have continuum-detected companions within the ALMA primary beam. We searched the spectra of any companions for emission lines at the same frequency as our detected line. AS2UDS.0109.1 has a tentative $5.3\sigma$ detection of an emission line which would correspond to {\Cii} at $z=4.451$. This would place it at the same redshift at AS2UDS.0109.0, however the line is very weak and the false positive rate at this significance and line width is $\sim50\%$. Further observations are therefore required to confirm the line. The spectrum and photometric thumbnails for AS2UDS.0109.1 are shown in Figure \ref{fig:suppl}. 

\section{Additional data and discussion of candidate {\Cii} emitters} \label{app:notes}

\begin{figure}  
\vspace{-0.cm} \noindent     

\includegraphics[width=\textwidth, trim={1.5cm 0.9cm 8cm 1cm}, clip]{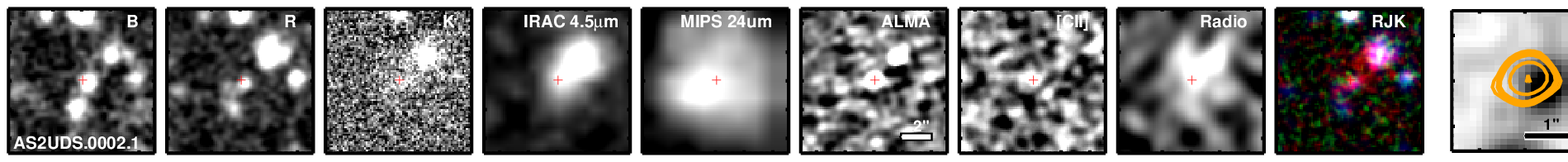}

\vspace{-1cm} \noindent     
\includegraphics[width=\textwidth, trim={1.5cm 0.9cm 8cm 1cm}, clip]{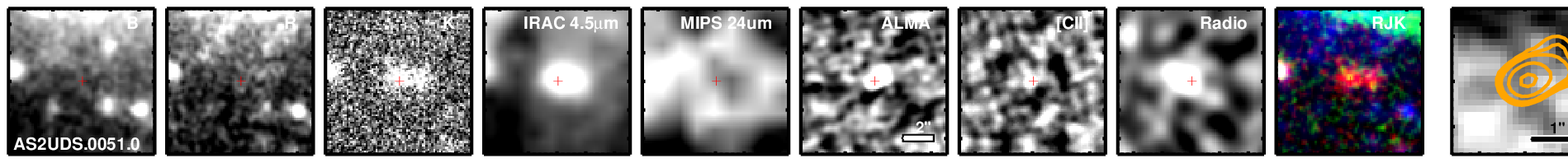}

\vspace{-1cm} \noindent     
\includegraphics[width=\textwidth, trim={1.5cm 0.9cm 8cm 1cm}, clip]{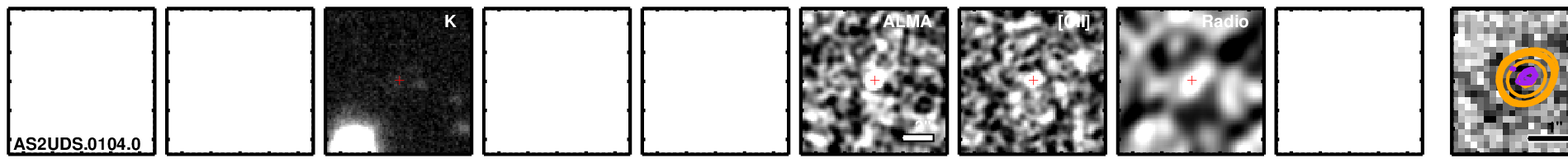}

\vspace{-1cm} \noindent     
\includegraphics[width=\textwidth, trim={1.5cm 0.9cm 8cm 1cm}, clip]{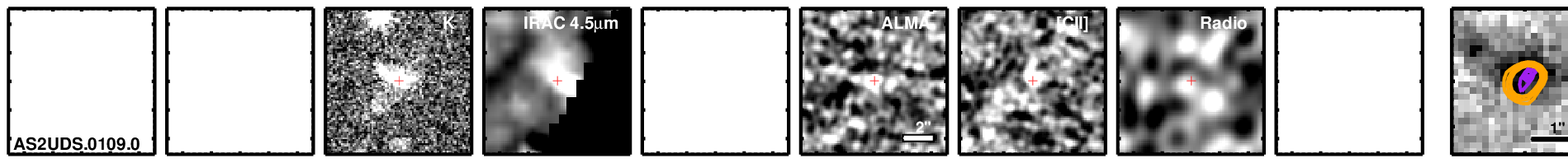}

\vspace{-1cm} \noindent     
\includegraphics[width=\textwidth, trim={1.5cm 0.9cm 8cm 1cm}, clip]{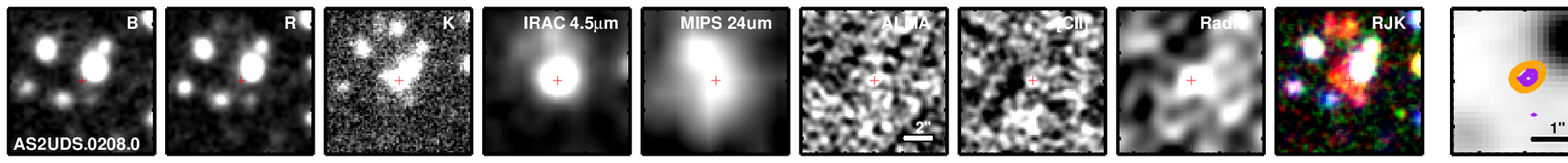}

\vspace{-1cm} \noindent     
\includegraphics[width=\textwidth, trim={1.5cm 0.9cm 8cm 1cm}, clip]{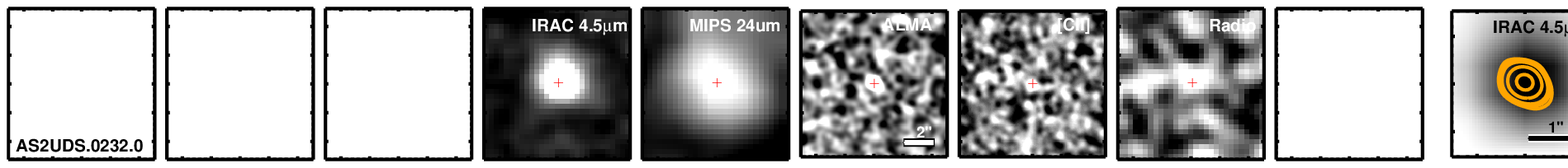}

\vspace{-1cm} \noindent     
\includegraphics[width=\textwidth, trim={1.5cm 0.9cm 8cm 1cm}, clip]{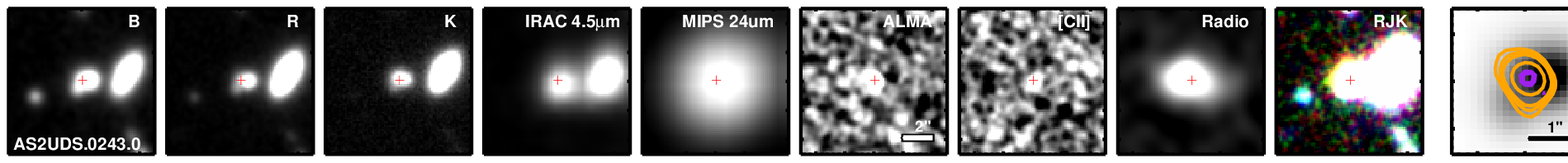}

\vspace{-1cm} \noindent     
\includegraphics[width=\textwidth, trim={1.5cm 0.9cm 8cm 1cm}, clip]{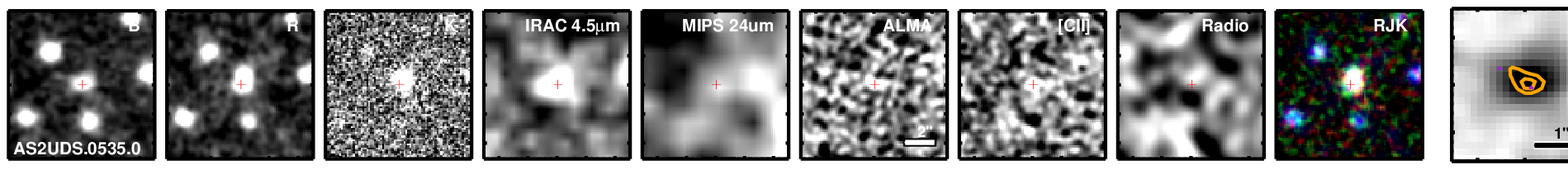}

\vspace{-1cm} \noindent     
\includegraphics[width=\textwidth, trim={1.5cm 0.9cm 8cm 1cm}, clip]{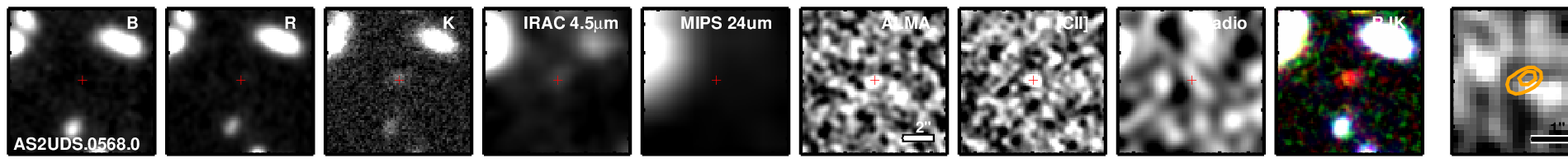}

\vspace{-1cm} \noindent     
\includegraphics[width=\textwidth, trim={1.5cm 0.9cm 8cm 1cm}, clip]{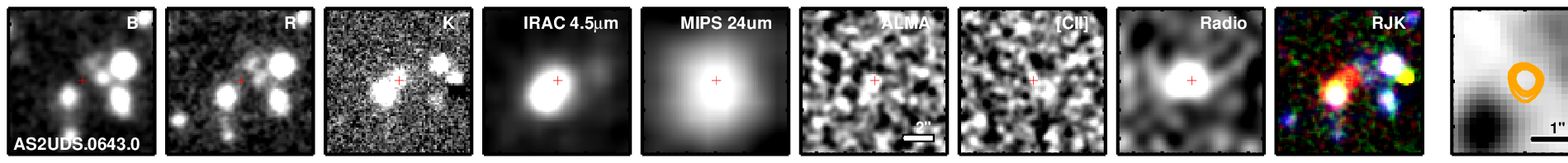}

\vspace{-0.5cm} \noindent 
\caption{
Multi-band thumbnails of main catalogue sources. The leftmost nine thumbnails are each $10''\times10''$, centred on the source submillimeter emission. The red cross marks the centre of the submillimeter emission from the ALMA maps. The rightmost thumbnail is a zoom $3''\times3''$ thumbnail with ALMA contours overlaid at $3, 4, 5, 10,$ and $20\sigma$. Purple contours are high-resolution ALMA continuum emission ($\sim0.15''$) where it is available, orange are at the tapered $0.5''$ resolution $870$\,\micron\ emission. 
}
\label{fig:allthumbs}
\end{figure}

Figure \ref{fig:allthumbs} shows $10''\times10''$ multiwavelength thumbnails, plus an enlarged view of the $B$-band and ALMA contours (see also Figure \ref{fig:thumbs}), for all ten line emitters in our sample. In this Appendix we discuss individual line-emitting sources in detail.

{\bf AS2UDS.0002.1} has a $B$-band detection within $1''$ of the $870$\,\micron\ emission, contaminating the photometry of the red component associated with the SMG. There is no $B$-band or radio emission at the position of the SMG, suggesting that the line we detect is likely to be {\CII} at $z\sim4.5$. However, this source may well be modestly lensed by the foreground galaxy.

{\bf AS2UDS.0051.0} has been confirmed with independent observations presented and discussed in detail in \citet{Gullberg2018}. 
There is no obvious optical/near-infrared counterpart in the multiwavelength thumbnails and so the most likely line identification is {\CII}. 

{\bf AS2UDS.0104.0} has the brightest emission line in our sample. There is no optical or \emph{Spitzer} coverage of this field so there is little photometric information available, however its $K$ band photometry and $1.4$\,GHz limit is consistent with a $z\sim4.5$ {\CII} emitter. 

{\bf AS2UDS.0109.0} is not covered by many optical or near-infrared images. It has a $K$ and $4.5$\,\micron\ detection, however it is undetected at $1.4$\,GHz and is therefore consistent with a $z\sim4.5$ {\CII} emitter.

{\bf AS2UDS.0208.0} lies within $1''$ of a bright $B$-band source with $z_{\rm phot}=2.5\pm0.2$ (Almaini et al., in preparation), suggesting this may be a lensed $z\sim4.5$ {\CII} emitter. The SMG itself does not have an optical counterpart, first appearing in the near-infrared and is undetected at $1.4$\,GHz. 

{\bf AS2UDS.0232.0} has no optical or near-infrared coverage. It is bright at $4.5$\,\micron\ and $24$\,\micron\ and undetected at $1.4$\,GHz. Its properties are consistent with the detected line being {\CII}. 

{\bf AS2UDS.0243.0}, in addition to a (slightly offset) $B$-band detection, is detected at 1.4\,GHz (with an offset between the ALMA and radio detections of $\sim0.25''$). It has an $870$\,\micron\ to 1.4\,GHz flux ratio consistent with a $z<4$ source ($\log_{10}(S_{870}/ S_{1.4})\sim0.5$). 
The photometric redshift derived for this source is $z=1.58\pm0.05$. This suggests the line we detect is CO($8$--$7$), giving a redshift of $z=1.63\pm0.01$ with a luminosity of $L_{\rm{CO}} = 1.5\times10^{8}$\,\Lsun.  
This source is therefore consistent with a $z=1.58$ CO($8$--$7$) line emitter.

{\bf AS2UDS.0535.0} has a $B$-band detection aligned with the $870$\,\micron\ emission. 
The photometric redshift derived for this source is $z=0.80\pm0.03$. If the photometry corresponded to the ALMA detection then the line we detect could be CO($5$--$4$) at $z=0.698\pm0.010$ with a luminosity of $L_{\rm{CO}} = 2.1\times10^{7}$\,\Lsun, or CO($6$--$5$) at $z=1.038\pm0.010$ with a luminosity of $L_{\rm{CO}} = 5.6\times10^{7}$\,\Lsun. However, this galaxy is undetected at mid-infrared and radio wavelengths, suggesting this is not a $z<2$ star-forming galaxy but rather a $z\sim4.5$ {\CII} emitter being lensed by a foreground galaxy. There is a secondary peak in the photometric redshift distribution at $z=4.63$, which may be more appropriate. 
 
{\bf AS2UDS.0568.0} has no detected optical counterpart (i.e., $B>28.4$, $R>27.7$) in the multi-wavelength thumbnails, suggesting this is a $z\sim4.5$ {\CII} emitter.  

{\bf AS2UDS.0643.0} has a photometric redshift of $z=4.44^{+0.62}_{-1.08}$. Albeit with large error bars, this is in agreement with the line we detect being {\CII} emission at $z=4.614\pm0.007$. There are multiple foreground sources with $z_{\rm phot}<4$ within $1$--$2''$ of AS2UDS.0643.0, which may be lensing the background SMG. This source is most likely to be a $z\sim4.6$ SMG. 

\newpage

\bibliographystyle{mn2e}
\bibliography{SMGs}


\end{document}